\documentclass[twocolumn,prd,aps,superscriptaddress,preprintnumbers,tightenlines,showpacs,nofootinbib,eqsecnum,amsfonts,amsmath]{revtex4}
\pdfoutput=1

\usepackage{color}
\usepackage{calc}
\usepackage{amsmath,amssymb,graphicx}
\usepackage{tensor}
\usepackage{bm}
\usepackage{times}
\usepackage[varg]{txfonts}
\usepackage{float}
\usepackage{dcolumn}

\usepackage[nolist,nohyperlinks]{acronym}
\usepackage{xspace}

\usepackage[abs]{overpic}
\usepackage{pict2e}

\usepackage[caption=false]{subfig} 

\DeclareMathAlphabet{\mathcalstd}{OMS}{cmsy}{m}{n}
\DeclareMathAlphabet{\mathpzc}{OT1}{pzc}{m}{it}

\newcommand{\UIB}{Departament de F\'isica, Universitat de les Illes Balears and Institut d'Estudis Espacials de Catalunya, 
Crta. Valldemossa km 7.5, E-07122 Palma, Spain}

\newcommand{\Cardiff}{School of Physics and Astronomy, Cardiff University, Queens Building, CF24 3AA, Cardiff, United Kingdom}
\newcommand{\ICTS}{International Centre for Theoretical Sciences, Tata Institute of Fundamental Research, IISc Campus, Bangalore 560012, India}
\newcommand{\AEI}{Max Planck Institute for Gravitational Physics (Albert Einstein Institute), Am M\"uhlenberg 1, Potsdam-Golm 14476, Germany}

\begin{document}

\preprint{LIGO-P1500143-v2}


\title{Frequency-domain gravitational waves from non-precessing black-hole binaries. \\ I. New numerical waveforms and anatomy of the signal}

\author{Sascha Husa}
\affiliation{\UIB}
\affiliation{\ICTS}

\author{Sebastian Khan}
\affiliation{\Cardiff}

\author{Mark Hannam}
\affiliation{\Cardiff}
\affiliation{\ICTS}

\author{Michael P\"urrer}
\affiliation{\Cardiff}

\author{Frank Ohme}
\affiliation{\Cardiff}

\author{Xisco Jim\'enez Forteza}
\affiliation{\UIB}

\author{Alejandro Boh\'e}
\affiliation{\UIB}
\affiliation{\AEI}

\begin{abstract}
In this paper we discuss the anatomy of frequency-domain gravitational-wave signals from non-precessing black-hole coalescences with the goal of constructing accurate phenomenological waveform models. We first present new numerical-relativity simulations for mass ratios up to 18, including spins. From a comparison of different post-Newtonian approximants with numerical-relativity data we select the uncalibrated SEOBNRv2 model as the most appropriate for the purpose of constructing hybrid post-Newtonian/numerical-relativity waveforms, and we discuss how we prepare time-domain and frequency-domain hybrid data sets. We then use
our data together with results in the literature to calibrate simple explicit expressions for the final spin and radiated energy. Equipped with our prediction for the final state we then develop a simple and accurate merger-ringdown-model based on modified Lorentzians in the gravitational wave amplitude and phase, and we discuss a simple method to represent the low frequency signal augmenting the TaylorF2 post-Newtonian approximant with terms corresponding to higher orders in the post-Newtonian expansion. We finally discuss different options for modelling the small intermediate frequency regime 
between inspiral and merger-ringdown. A complete phenomenological model based on the present work is presented in a companion paper \cite{Khan2015}.
\end{abstract}


\maketitle

\acrodef{PN}{post-Newtonian}
\acrodef{EOB}{effective-one-body}
\acrodef{NR}{numerical relativity}
\acrodef{GW}{gravitational-wave}
\acrodef{BBH}{binary black hole}
\acrodef{BH}{black hole}
\acrodef{BNS}{binary neutron star}
\acrodef{NSBH}{neutron star-black hole}
\acrodef{SNR}{signal-to-noise ratio}
\acrodef{aLIGO}{Advanced LIGO}
\acrodef{AdV}{Advanced Virgo}

\newcommand{\PN}[0]{\ac{PN}\xspace}
\newcommand{\EOB}[0]{\ac{EOB}\xspace}
\newcommand{\NR}[0]{\ac{NR}\xspace}
\newcommand{\BBH}[0]{\ac{BBH}\xspace}
\newcommand{\BH}[0]{\ac{BH}\xspace}
\newcommand{\BNS}[0]{\ac{BNS}\xspace}
\newcommand{\NSBH}[0]{\ac{NSBH}\xspace}
\newcommand{\GW}[0]{\ac{GW}\xspace}
\newcommand{\SNR}[0]{\ac{SNR}\xspace}
\newcommand{\aLIGO}[0]{\ac{aLIGO}\xspace}
\newcommand{\AdV}[0]{\ac{AdV}\xspace}

\section{Introduction}
\label{sec:introduction}

Our goal is to develop an accurate and simple description of the \GW signal of a
compact binary coalescence in the form of explicit expressions in the
frequency domain, which are particularly convenient for \GW data analysis.
Optimal methods for the detection and accurate identification of events are
based on filtering the data stream with a template bank of physically correct
waveforms. Extracting the maximal amount of information from the data thus
relies critically on the availability of accurate waveform models, such as the
ones we discuss here. For small masses, such as for neutron-star binaries, the
\PN description \cite{Blanchet2014} is considered sufficient for
searches for \GW events in the advanced detector era
\cite{Ajith:2007xh,Buonanno2009}. However, as the binary mass increases and the
actual merger comes into band, the \PN approximation breaks down and
non-perturbative solutions for the Einstein equations are required. For the
simple case of non-spinning \acp{BH}, a comparison of different
\PN results suggests that such non-perturbative information is
required for the efficient detection of binaries with a total mass larger than
12 solar masses \cite{Buonanno2009}. In this work we will treat the case of
coalescing BHs, or neutron stars spiralling into sufficiently heavy BHs such
that equation of state effects can be neglected. The final state of the
coalescence, a single perturbed Kerr BH, can also be treated with linear
perturbation theory. This allows us to compute the complex frequencies of damped
sinusoid spheroidal harmonic  ``quasinormal'' modes, but not their amplitudes or
relative phase.

In the present paper we focus on understanding the anatomy of frequency-domain
waveforms and laying the groundwork for accurate frequency-domain inspiral-merger-ringdown (IMR) waveform
models, including a discussion of how we combine information from numerical
relativity and perturbative approaches. In a second paper \cite{Khan2015} we
describe in detail the construction of a  specific model, ``PhenomD'', which we
have implemented in the LIGO Algorithms Library (LAL)~\cite{lalsuite}, the key
software infrastructure for \GW data analysis from interferometric ground based
detectors, and we compare with other waveform models implemented in LAL, in
particular SEOBNRv2 \cite{Taracchini2014}. While the main motivation is to
provide a practical tool for \GW astronomy that we hope to be useful for
searches or parameter estimation, we stress that gaining a better understanding 
of waveforms is by itself valuable for \GW physics.
 
Astrophysical models of compact binary merger event rates lead to the
expectation of a first direct detection of GWs within the next five years
\cite{Aasi2013,Abadie:2010cf}, as a new generation of interferometric detectors
is approaching design sensitivity 
\cite{Abbott:2007kv, 2010CQGra..27h4006H, advLIGO,Accadia:2011zzc}, expanding
the volume of the universe observable in the \GW window by roughly a factor of
1000 over previous detectors. This increase is expected to be sufficient to
exhaust the current large uncertainties in event rates, which are connected to
the large uncertainties about the populations of compact binaries in the
universe. Our lack of understanding of binary populations calls for being prepared
to cover as much as possible of the physical \BBH parameter space, and the
waveform model we discuss has been calibrated to the largest region of the
non-precessing parameter space to date, covering mass ratios up to 18.

The first successful calculation of the \GW signal during roughly the last
orbit, merger and subsequent quasi-normal-mode (QNM) ringdown
from a numerical solution of the Einstein equations, without perturbative
approximations, dates back only one decade \cite{Pretorius2005}. Since then,
synthesizing models of the complete coalescence waveform from \PN
results, \NR and \BH perturbation theory has become a
key goal of \GW source modeling.
One approach has been to model gravitational waveforms in the time domain, by
calibrating an \EOB resummation
\cite{Buonanno:1998gg,Buonanno:2000ef} of \PN results to \NR
simulations
\cite{Buonanno:2007pf,Damour:2008te,Pan:2011gk,Damour:2012ky,Taracchini2014,
Nagar:2015xqa}. This approach provides very accurate waveforms, but still
requires the solution of a system of ordinary differential equations for each
case and is computationally expensive, in particular for applications in
parameter estimation and model selection within a Bayesian framework. A route to
speed up the evaluation of \EOB  models by means of reduced order techniques is
discussed in \cite{Field:2013cfa,Puerrer2014,Puerrer2015inprep}.

Our previous work has followed an alternative approach, where we have modeled
the wave signal with explicit expressions in the frequency domain. This approach
offers simplicity and evaluation speed, however previous models showed
relatively poor extrapolation beyond their calibration region and limited
accuracy for parameter estimation. In this work we redesign our phenomenological
approach based on a more detailed study of the anatomy of waveforms in the
frequency domain and calibrate the model with a larger and more accurate set of
numerical waveforms up to a mass ratio of 18.  As for previous non-precessing
phenomenological waveform models
\cite{Ajith2007,Ajith:2007kx,Ajith2011,Santamaria2010}, this work is restricted
to  the dominant $\ell=|m|=2$ spherical harmonic mode, leaving the general
aligned spin case, higher harmonics and precession for future work. From the
point of view of \GW data analysis, the model we construct here can directly
replace our previous PhenomB and PhenomC waveform models with similar
computational cost but an accuracy that for a large part of the parameter space
further improves upon current \EOB-NR models. A route toward extending such
models to precession has been described in \cite{Hannam:2013oca}. 

In the next section, we will discuss our ``input'' waveforms:
\NR waveforms, inspiral waveforms computed in the
\EOB description, and other \PN approximants, and how we construct
complete time domain hybrid waveforms.  Our approach does in fact allow us to model
the waveforms in pieces, e.g. to calibrate the inspiral and merger-ringdown
models separately, without ever constructing a hybrid first. However, the
construction of complete time domain hybrids is convenient for comparisons, and
to develop our modeling approach without artificial restrictions due to the
length of available waveforms. At the hybridisation stage we will compare the
agreement of different \PN approximants with our numerical data, and choose to
hybridize with ``uncalibrated'' SEOBNRv2 waveforms, i.e., where we have removed
all \NR calibrations. 

In Sec. \ref{sec:finalstate} we will refocus from the inspiral to the modelling of the final state and present fits for the final spin and radiated energy as functions of mass ratio and spin. These fits will provide crucial information regarding the modelling of the ringdown. With the final state information and a complete time domain hybrid in hand, we will turn to the study of the anatomy of waveforms in the Fourier domain which will motivate our plan for construction of a waveform model, presented in Sec. \ref{sec:model}. We conclude with a summary in Sec. \ref{sec:conclusions}.


\section{Input waveforms and hybrid construction}
\label{sec:input}

\subsection{Waveform notation and conventions}

Our work concerns the $\ell=\vert m\vert = 2$ spherical harmonic modes of the
complex \GW strain $h_{22}$, which we decompose into an amplitude $A$ and phase
$\phi$, which depend on time $t$, and the intrinsic parameters of the source
$\Xi$,
\begin{equation}
h_{22}(t, \Xi)=A(t,\Xi)e^{-i\phi(t,\Xi)}.
\end{equation}
The intrinsic parameters $\Xi$ consist of the dimensionless projections of the BH spins $\vec S_{1,2}$ in the direction of the orbital angular momentum $\hat L$, and the masses $m_{1,2}$, where
\begin{equation}
\chi_i = \frac{\vec S_i \cdot \vec L}{m_i^2 \, \vert \vec L \vert}, 
\end{equation}
and we define the mass ratio $q = m_1/m_2 \geq 1$, total mass $M=m_1+m_2$ and symmetric mass ratio $\eta = m_1 m_2 /M^2$.
Our binaries are non-precessing, and as usual we choose our coordinates and spherical harmonic modes consistent with the equatorial symmetry exhibited by such spacetimes, such that
\begin{equation}
\label{eq:alignedspinsymmetryhlm}
h_{22} = h^{*}_{2,-2}.
\end{equation}
We also  assume negligible eccentricity, in which case the frequency
$
\omega(t)={d \phi}/{dt}
$
is a monotonic function of $t$.  Residual eccentricity in numerical waveforms
and potentially other numerical artefacts will cause oscillations, however,
provided these are small enough, monotonicity is preserved until merger, and we can define
the inverse function $t(\omega)$.

We define the Fourier transform $\tilde h(\omega)$ of a complex function $h(t)$ as
\begin{equation}\label{eq:def:Fourier}
\tilde h(f) = \int_{-\infty}^{\infty} h(t) e^{-i 2 \pi f \, t} \, dt,
\end{equation}
consistent with the conventions adopted in the LIGO Algorithms Library
\cite{lalsuite}. We will also use the radial frequency $\omega = 2
\pi f$, in particular in quantitative statements, to 
facilitate comparisons with data analysis considerations. With our convention of the Fourier transform, time derivatives
are converted to multiplication in the Fourier domain by $i 2 \pi f$. 
We recall, that as a consequence of time derivatives being converted to multiplication in
Fourier space, the standard conversion between \GW strain and the Newman-Penrose
scalar $\psi_4$, where
\begin{equation}
\frac{d^2 h(t)}{dt^2} = \psi_4(t),
\end{equation}
only affects the Fourier domain amplitude, but not the phase, up to a jump of $\pi$, and apart from possible effects
specific to the numerical algorithm used to carry out this conversion.

For \NR data or hybrid waveforms we need to Fourier transform discrete time
series $h_r$, then the above convention implies
\begin{equation}\label{eq:defDiscreteFourier}
\tilde h_s = \Delta t \sum_ {r = 0}^{n-1}  h_r e^{- 2\pi \, i \, r \, s/n},
\end{equation}
where $\Delta t$ is the uniform time step of the time series.

The low frequency limit for the Fourier domain waveform can be understood from \PN theory. Analytically
Fourier-transforming the TaylorT2 approximant with the stationary-phase-approximation
yields the TaylorF2 approximant, an explicit expression of the Fourier domain strain in the Frequency domain.
To leading order the amplitude diverges at low frequency as $f^{-7/6}$, and the phase as $f^{-5/3}$, see the discussion in Secs.~\ref{sec:amp} and \ref{sec:phase} below. 

For the modelling of IMR waveforms we also need to understand the falloff properties of the waveform at high frequencies.
For smooth square integrable functions of time it is well known (see e.g.~\cite{TrefethenWeb1996}) that
\begin{equation}
\tilde h(f) = O(\vert  f \vert^{-M}) \quad as \quad \vert f\vert \rightarrow \infty \quad\mbox{for all}\, M,
\end{equation}
thus the Fourier transform then falls off faster than any power law at high frequency. 
If a function $h(t)$ only has $p$ continuous derivatives in $L^2$ for some $p\geq 0$  and a $p$th derivative in $L^2$ of bounded variation, then \cite{TrefethenWeb1996}
\begin{equation}\label{eq:FourierFalloff}
\tilde h(f) = O(\vert  f \vert^{-p-1}) \quad as \quad \vert f\vert \rightarrow \infty.
\end{equation}
While the Fourier domain strain diverges as $f^{-7/6}$ at low frequencies, and is thus not square integrable for idealized infinitely long signals (in contrast to finite astrophysical signals), the time derivative of the strain (the news function) is square integrable (being proportional to $f^{-1/6}$ at low frequencies), and the faster-than-power-law falloff for the strain can be inferred from the scaling of the Fourier amplitude when taking time derivatives.

\subsection{Numerical relativity waveforms}\label{sec:NR}

Our numerical waveforms have been taken from two independent data sets, the publicly available SXS catalogue \cite{SXS:catalog} (computed by the SpEC code~\cite{Ossokine:2013zga,Hemberger:2012jz,Szilagyi:2009qz,Scheel:2008rj,Boyle:2007ft,Mroue:2013xna, Buchman:2012dw}), and from a set of waveforms that have recently been constructed with the BAM code \cite{Bruegmann2008,Husa2008}, and which are listed in tables \ref{tab:BAMwfs}, \ref{tab:BAMwfs2}. Both the SXS and BAM data sets are divided into waveforms we use to calibrate the PhenomD model, and a larger set we use for model verification. 
The key data sets that determine the calibration range of our waveform model are
the BAM waveforms for a range of spins at mass ratio 1:18, high-spin BAM data at mass ratios 4 and 8, and 
equal mass SXS data sets at very high spins of $-0.95$ and $+0.98$. 

\begin{table}
\begin{tabular}{ lccccccr}
\hline
\hline
$q$  & $\chi_1$ & $\chi_2$ & $D/M$ & $p_t/M$ & $- p_r/M$ & $M\omega_i$ & $N_{\rm GW}$ \\ 
         &              &                &              &              & $\ (\times 10^{-4})$ &         & \\
\hline
 2 &  0.50 &  0.50 &  12.2 &  0.07277 &  3.468 &  0.04128 &  26.8  \\
 2 & 0.75 & 0.75 & 11.1 & 0.07591 & 3.404 & 0.04689 & 23.5  \\
 3 & -0.50 & -0.50 & 12.8 & 0.06309 & 2.870 & 0.03958 & 19.9  \\
 4 & -0.75 & -0.75 & 12.1 & 0.05694 & 1.858 & 0.04283 & 15.6  \\
 4 & -0.50 & -0.50 & 11.9 & 0.05669 & 2.694 & 0.04352 & 18.1  \\
 4 & -0.25 & -0.25 & 11.5 & 0.05694 & 2.784 & 0.04528 & 18.9  \\
 4 & 0.00 & 0.00 & 11.2 & 0.05710 & 2.829 & 0.04673 & 21.1  \\
 4 & 0.25 & 0.25 & 11.0 & 0.05681 & 2.962 & 0.04785 & 23.0  \\
 4 & 0.50 & 0.50 & 10.8 & 0.05648 & 4.493 & 0.04870 & 26.1  \\
 4 & 0.75 & 0.75 & 10.7 & 0.05604 & 1.136 & 0.04931 & 30.0  \\
  8 & -0.85 & -0.85 & 10.0  & 0.04135 & 2.6142 & 0.06486 & 8.2  \\
 8 & 0.80 & 0.00 & 8.00 & 0.04146 & 3.009 & 0.07266 & 23.3  \\
  8 & 0.85 & 0.85 & 6.50 & 0.04703 & 8.0706 & 0.10951 & 15.7  \\
 10 & 0.00 & 0.00 & 8.39 & 0.03670 & 1.685 & 0.06942 & 13.4  \\
 18 & -0.80 & 0.00 & 10.0 & 0.02080 & 0.6589 & 0.05614 & 14.3  \\
 18 & -0.40 & 0.00 & 9.00 & 0.02176 & 0.7842 & 0.06395 & 15.1  \\
 18 & 0.00 & 0.00 & 7.58 & 0.02379 & 1.261 & 0.07932 & 13.0  \\
 18 & 0.40 & 0.00 & 7.43 & 0.02300 & 1.161 & 0.08052 & 23.2  \\
 \hline
 \hline
\end{tabular}
\caption{\label{tab:BAMwfs}
Details of new BAM simulations. For each mass ratio $q$ and choice of spins $\chi_1$ and $\chi_2$, the initial binary separation
is $D/M$, and the tangential and orbital momenta are $(p_t,p_r)/M$. The initial \GW frequency is $M\omega_i$, and the
simulation completes $N_{\rm GW}$ orbits before merger.}
\end{table}

\begin{table}
\begin{tabular}{ lccccr}
\hline
\hline
$q$  & $\chi_1$ & $\chi_2$ & eccentricity                             & $ M_f $ & $ a_f$ \\ 
        &               &                &      $\ (\times 10^{-3})$ &             & \\
\hline
 2& 0.50& 0.50& 1.2& 0.945& 0.805\\
 2& 0.75& 0.75& 4.4& 0.930& 0.889\\
 3& -0.50& -0.50& 1.0& 0.9779& 0.300\\
 4& -0.75& -0.75& 0.8& 0.9846& 0.049\\
 4& -0.50& -0.50& 1.0& 0.9831& 0.194\\
 4& -0.25& -0.25& 1.0& 0.981& 0.334\\
 4& 0.00& 0.00& 1.4& 0.978& 0.472\\
 4& 0.25& 0.25& 2.4& 0.9738& 0.607\\
 4& 0.5& 0.5& 3.6& 0.9674& 0.738\\
 4& 0.75& 0.75& 4.0& 0.9573& 0.863\\
 8& -0.85& -0.85& 0.5& 0.9931& -0.320\\
 8& 0.80& 0.00& 4.9& 0.977& 0.860\\
 8& 0.85& 0.85& 9.1& 0.9746& 0.895\\
 10& 0.00& 0.00& 0.8& 0.9918& 0.255\\
 18& -0.80& 0.00& 0.5& 0.9966& -0.531\\
 18& -0.40& 0.00& 0.5& 0.9966& -0.188\\
 18& 0.00& 0.00& 1.3& 0.9959& 0.163\\
 18& 0.40& 0.00& 1.8& 0.9943& 0.505\\
 \hline
 \hline
\end{tabular}
\caption{\label{tab:BAMwfs2}
Eccentricity measured from the orbital frequency, final Kerr parameter and final mass of the new BAM simulations.}
\end{table}

\begin{table}
\begin{tabular}{lllllllll}
  \hline
  \hline
  \text{Code/ID}  & \text{$\chi_1$} & \text{$\chi_2$} &
  $M_f$ & $a_f$ & $Mf_{\text{RD}}$ & $Mf_{\text{hyb}}$\\
  \hline
 \text{SXS:BBH:0001}  & 0.00 & 0.00 &  0.9516 & 0.6865 & 0.0881
 & 0.00398 \\
 \text{SXS:BBH:0156}  & -0.95 & -0.95 & 0.9681 & 0.3757
& 0.0713 & 0.00522 \\
 \text{SXS:BBH:0172}  & 0.98 & 0.98 & 0.8892 & 0.9470 &
0.1328 & 0.00497 \\
\hline
 \hline
\end{tabular}
\caption{
High spin equal mass waveforms from the SXS catalogues used in this paper to illustrate waveform anatomy. For each
configuration we list the SXS catalogue number,  the spins
$\chi_1$ and $\chi_2$. The
final \BH has mass $M_f$ and 
    dimensionless spin $a_f$, and the QNM ringdown signal has frequency $Mf_{\rm RD}$. 
    The frequency $Mf_{\text{hyb}}$ marks the midpoint of the transition region between
 SEOBv2 inspiral and \NR data.}
 \label{tab:SXSwftable} 
 \end{table}

SpEC is a multi-domain pseudo-spectral code that uses
excision techniques to remove the \BH interiors (and, most importantly, each
\BH singularity)
from the computational domain. For an in-depth explanation see 
Refs.~\cite{Scheel2014, Szilagyi2014, SXS:catalog}.
The SpEC code evolves the Einstein field equations using
the Generalised Harmonic coordinate formulation. This code has been used to 
produce the longest simulations to date (175 orbits / 350 \GW cycles for a
mass-ratio 1:7 
configuration~\cite{Szilagyi:2015rwa}), 
and 201 waveforms have been made publicly available as of June 
2015~\cite{SXS:catalog}; see Ref.~\cite{Mroue:2013xna} for details. 

A key set of waveforms in the SXS catalogue are those for highly spinning equal-mass
\BH binaries. Most codes in the field (including BAM and SpEC) primarily use 
conformally flat initial data, which limits the \BH spins to 
$a/m \sim 0.92$~\cite{Cook:1989fb,Lovelace:2008tw,Hannam:2009ib}.  
There has been some work in constructing high-spin data for puncture codes like 
BAM~\cite{Hannam:2006zt,Ruchlin:2014zva}, 
but to date no waveforms for long-term quasi-circular inspirals have been published. 
For the SpEC code, however, high-spin data \emph{have} been produced and used in production
simulations, based on a superposition of Kerr-Schild data~\cite{Lovelace:2008tw,Lovelace:2011nu,Mroue:2013xna}. 
This technique has made possible a number of equal-mass
binary simulations with high spins, and these have been used here both in the calibration of our 
model, and in tests of its accuracy and fidelity. 

The SXS waveforms that we use all cover 12 orbits or more of inspiral before merger, and the waveforms
are extrapolated to null infinity at third order as discussed in \cite{Boyle:2009vi}.
In this paper we will use 3 waveforms as listed in Table \ref{tab:SXSwftable}
from the SXS catalogue to illustrate waveform anatomy, and for comparison with
different \PN approximants.

The BAM code \cite{Bruegmann2008, Husa2008} solves the 3$+$1 decomposed
Einstein evolution equations using the $\chi$-variant of the
moving-punctures version of the BSSN \cite{Shibata:1995, Baumgarte:1998}
formulation. Spatial derivatives are sixth-order accurate in the bulk
\cite{Husa2008}. Kreiss-Oliger dissipation terms converge at fifth order,
and a fourth-order Runge-Kutta algorithm is used for the time evolution.
\BBH puncture initial data \cite{Brandt1997, Bowen1980} are calculated
with the pseudo-spectral elliptic solver described in Ref.~\cite{Ansorg2004}).
The \acp{GW} are calculated using the Newman-Penrose scalar
$\Psi_{4}$. For further details see Ref.~\cite{Bruegmann2008}.
The \acp{GW} are extracted at a finite distance, typically $\sim100M$ from the
source. 

For this work, new simulations were performed at mass ratios 2, 3, 4, 8 and 18.
We also repeat the nonspinning 1:10 simulation reported in Refs.~\cite{Lousto:2010qx,Nakano2011},
using the same parameters as given in that work. 

For the simulations at mass ratios $\leq 4$, the eccentricity was reduced to $e \lesssim 10^{-3}$ using
the procedure described in Ref.~\cite{Puerrer2012}, which is an iterative procedure based on 
comparisons against \PN and \EOB  inspirals. At mass ratios 8 and 18, some of the
simulations were too short to reliably use this procedure. However, for large mass ratios the time scale of 
significant early time perturbations of the orbital quantities due to gauge transients is smaller than for comparable masses, which allows to easily relate observed residual eccentricity in the orbital motion to an excess or deficiency in the value of the initial tangential component of the black hole momentum, which dominates the error with respect to the true quasi-circular inspiral parameters. Consequently, a straightforward iteration ``by hand'' can be performed rather easily, starting with parameters calculated from \PN inspirals.

The length of our \NR waveforms, summarized in tables \ref{tab:BAMwfs}, \ref{tab:BAMwfs2}, was chosen
partly to sensibly connect them to inspiral approximants. \PN inspiral waveforms
become less accurate at higher frequencies, and we need to decide at what
frequency we need to be able to switch to \NR information in order to produce a
sufficiently accurate model. Previous work suggests that $\sim$5--10 orbits will
be sufficient to produce models that meet the needs for \GW detections with
advanced \GW detectors \cite{Hannam:2010ky,Ohme:2011zm}, if we use standard \PN
approximants for the inspiral. Those works took the largest differences between
the highest-order \PN waveforms available at that time as a conservative
estimate of their overall error. Here, however, we push the boundaries of the
\NR-accessible parameter space to higher mass ratios and spins, and we aim to
produce a waveform model suitable not only for detection but also
parameter-estimation purposes in the advanced detector era. 

If we were to follow the stringent accuracy requirements for combining (any) \PN
approximant with \NR data, we would need to simulate hundreds of \NR orbits
\cite{MacDonald:2011ne,MacDonald:2012mp,Damour:2010zb,Boyle:2011dy}, which is
prohibitively expensive, especially given the parameter space that we are
covering. At mass ratio 18, medium-resolution simulations
of only 8 orbits required 800000 CPU hours and high-resolution simulations 1.3
million CPU hours. Previous length-requirement studies, however, conservatively
treated all \PN approximants as equally (in)accurate, and it is worth
investigating whether an individual inspiral description is more accurate than
others, thereby allowing a matching with \NR data at higher frequencies.
While this question is difficult to answer exhaustively without extremely long
\NR waveforms, we discuss in Sec.~\ref{sec:hybrids} that the dramatically
better agreement between \NR and members of the \EOB family complements the
evidence that \EOB is a more accurate inspiral description
\cite{Boyle:2008ge,Szilagyi:2015rwa,Kumar:2015tha}, hence we can connect it reliably to our
reasonably short high-mass-ratio simulations.

We must also ensure that our simulations are sufficiently accurate. In the past we have performed detailed
convergence tests of our code~\cite{Bruegmann2008}, which provide strong evidence that the numerics are correct. 
Since performing the simulations reported in Ref.~\cite{Hannam:2010ec}, we have found that the small number of 
AMR buffer zones (six) limited the accuracy in long unequal-mass simulations. This is the cause of the 
disagreement between the BAM and SpEC nonspinning $q=4$ simulations that were discussed in 
Sec.~IV.A of Ref.~\cite{MacDonald:2012mp}. In our current simulations we use a minimum of 16 buffer
zones, which is sufficient to buffer mesh refinement boundary effects during a fourth-order Runge-Kutta time step of a fine grid
for sixth order finite differencing and compatible Kreiss-Oliger dissipation \cite{Kreiss1973}.

\begin{figure}[htbp]
\includegraphics[width=\columnwidth]{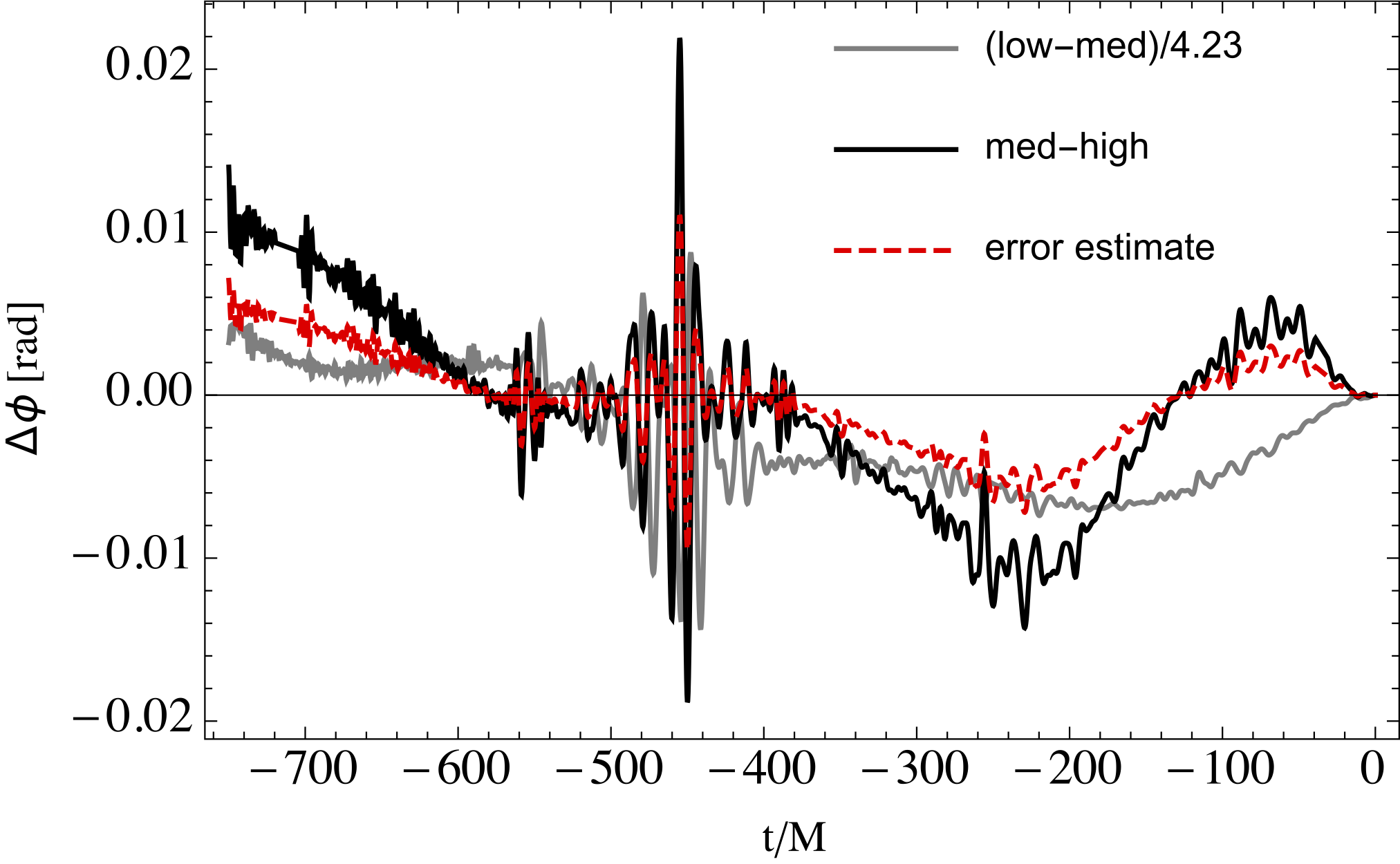}  
\caption{Convergence and Richardson-extrapolated error for non-spinning BAM simulations at $q=18$ at resolutions  $(96,120,144)$  as described in the main text.}
\label{fig:q18:conv}
\end{figure}

\begin{figure}[htbp]
\includegraphics[width=\columnwidth]{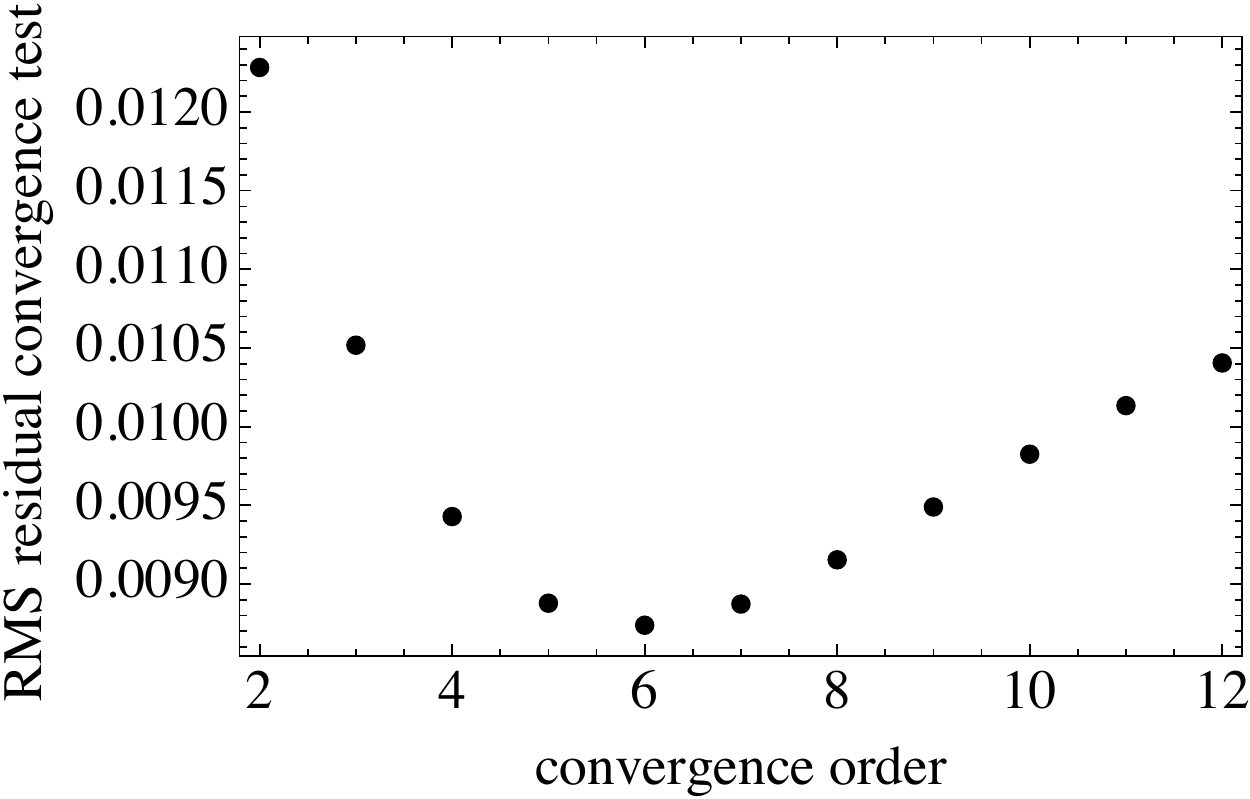} 
\caption{RMS difference between different resolutions as defined in Eq.~(\ref{eq:defRMS_Convergence})  for non-spinning BAM simulations at $q=18$ at resolutions  $(96,120,144)$  as described in the main text. Rescaling for 6th order convergence as shown in Fig.~ \ref{fig:q18:conv} is most consistent with our data.
}
\label{fig:q18:convRMS}
\end{figure}

\begin{figure*}[hbtp]
\includegraphics[width=\columnwidth]{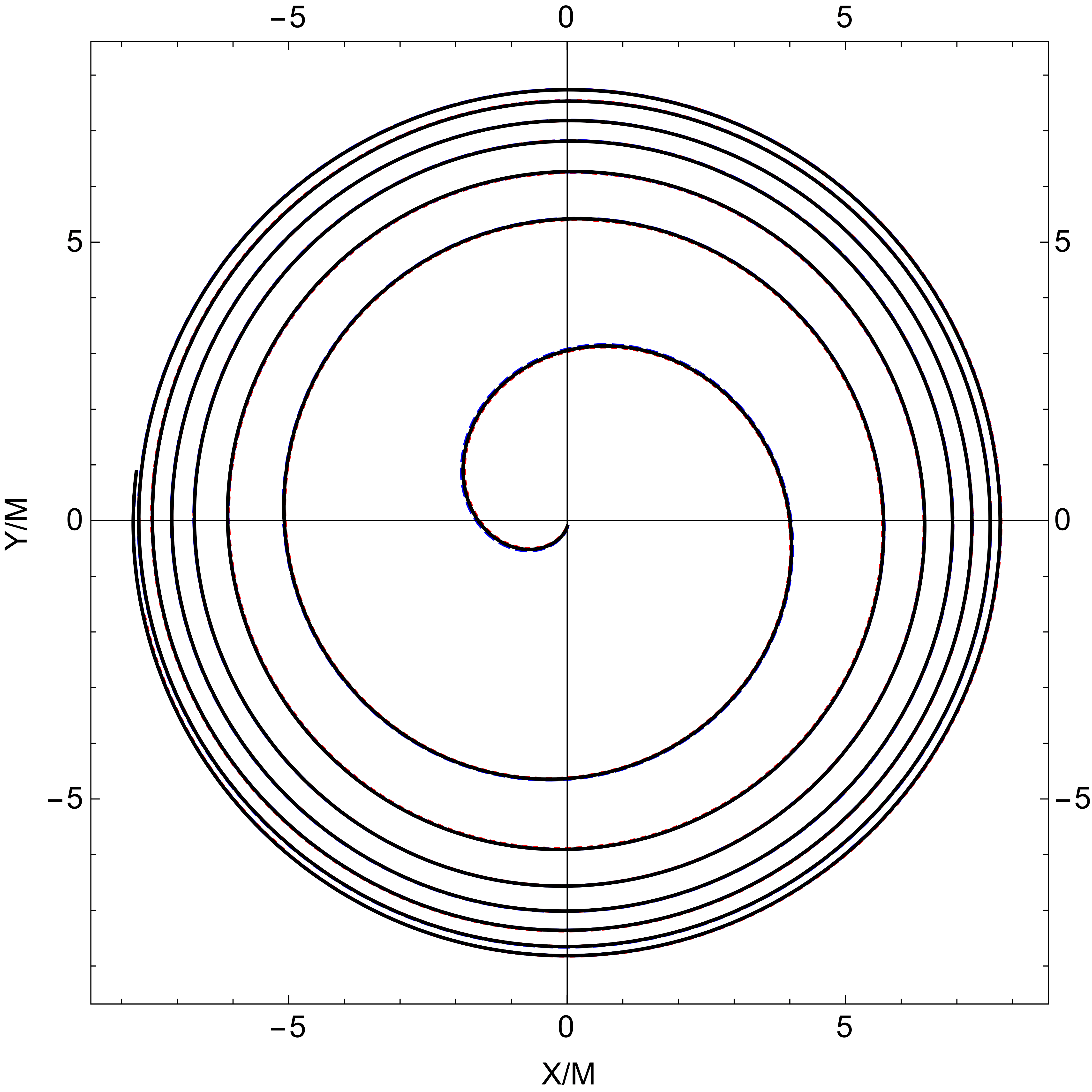}    
\includegraphics[width=\columnwidth]{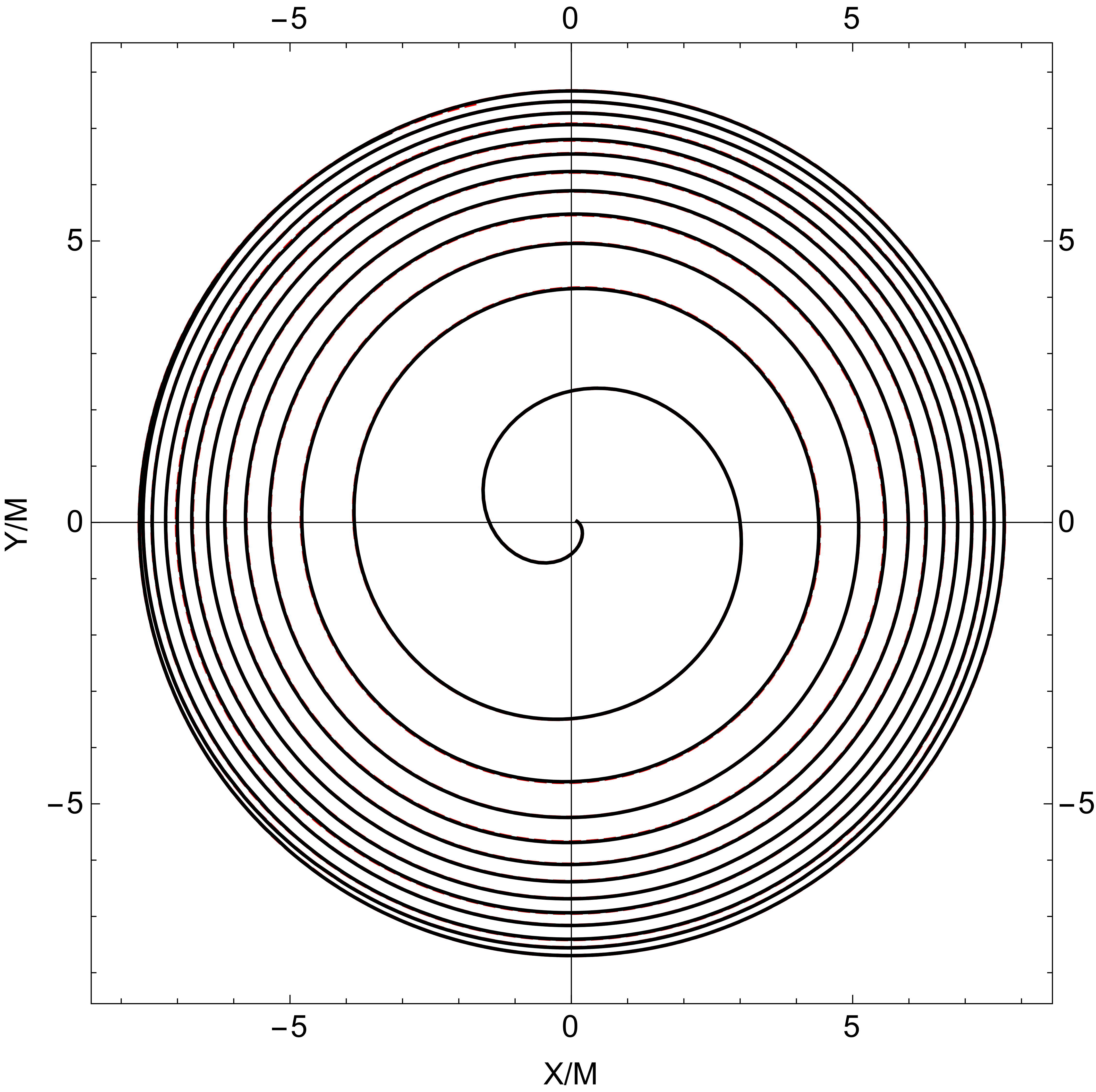}  
\caption{
Left: Orbital motion of 3 highest resolutions (96, 120, 144) for non-spinning $q=18$ case, aligned at aligned at $M \omega_{orb}=0.2$. 
Right:  Orbital motion of 2 highest resolutions (96, 120) for  $q=18$, $\chi_{1,2} = 0.4, 0$ case, aligned at $M \omega_{orb}=0.1$.
}
\label{fig:q18:orbits}
\end{figure*}

Even for the lowest-cost configuration (an equal-mass, non spinning binary), the computational cost 
prohibits a standard factor-of-two convergence test, and for many configurations there have been no
clean demonstrations of convergence for \emph{any} \BBH code. Our
confidence in \NR results
in general is based on the strong agreement between simulations from many independent codes, e.g., 
Refs.~\cite{Hannam:2009hh,Ajith:2012az,Hinder:2013oqa}. The results of a
convergence test for a series of simulations for the $q=18$ nonspinning case are
shown in Fig.~{\ref{fig:q18:conv}. Due most likely to poorly resolved initial
gauge dynamics and an initial burst of unphysical \GW content (``junk
radiation''), we have not been able to show convergence of the phase aligning at
the beginning of the simulation, but we see reasonable sixth order convergence
when aligning the $\ell=m=2$ \GW signal near the merger at a frequency of 
$M \omega_{GW} = 0.2$.  
Fig.~{\ref{fig:q18:conv} shows results from grid configurations with $(96,120,144)$ points in the innermost mesh refinement boxes, which dominate phasing errors and computational cost. Differences are scaled to sixth order convergence and we also show the error obtained as the difference of the highest resolution (144) and the Richardson-extrapolated result. Secular dephasing is very small, below $0.04 rad$, and errors are dominated by oscillations and noise. Due to the oscillations and noise it is hard  to decide by eye which convergence order is most consistent with the data, so we considered the RMS deviation
\begin{equation}\label{eq:defRMS_Convergence}
RMS^2 = \frac{1}{t_2-t_1}\int_{t_1}^{t_2} \left(  (X_1 - X_2)/C(n) - (X_2 - X_3) \right)^2 \, dt,
\end{equation} 
where the $X_i$ are the phases at resolutions $(96,120,144)$ aligned at $M \omega_{GW} = 0.2$, $C$ is the rescaling factor corresponding to convergence order $n$, $t_1$ and $t_2$ are the minimal and maximal times shown in Fig.~\ref{fig:q18:conv}, and the unresolved pulse of noise between times $-500$ and $-400$ is set to zero. As can be seen in Fig.~\ref{fig:q18:convRMS}, the best-fitting convergence order is 6.
In addition, in Fig.\ref{fig:q18:orbits} we show the that the orbital motion for the nonspinning convergence series and a low and medium resolution for a spinning $q=18$ case is practically indistinguishable for different resolutions. We have therefore not carried out a higher resolution run for the spinning case to save another $\approx 2$ million CPU hours, and we use the $96$ points resolution or better for the $q=18$ production runs.

Our \NR waveforms only model the waveform from an astrophysical inspiral after
some time $t_0$, when the influence of initial transients can be neglected. 
When processing a large number of waveforms, it is desirable to automatize the tedious procedure to find appropriate choices of $t_0$, and we thus need to understand initial transients in the waveform well enough to exclude them with a general algorithm. 
Two types of transients are of particular interest: the unphysical radiation
content in the initial data, often referred to as ``junk radiation'', and a
transient component resulting from our conversion of $\psi_4$ to strain. In
order to estimate the time on which junk radiation degrades the quality of the
wave signal, we first determine the maximum of the amplitude of the junk
radiation $t_J$ for the \NR $\psi_4$ waveforms, and then add a multiple of the
QNM damping period $T_{RD} = 1/f_{damp}$ corresponding to the two initial BHs:
\begin{equation}\label{eq:cleanTime}
\Delta T_J = t_J + \sigma_J T_{RD}.   
\end{equation}
In order to convert $\psi_4$ waveforms to strain we use the fixed frequency integration algorithm \cite{Reisswig:2010di}, 
which depends on the choice of cutoff frequency $f_0$. We choose this frequency
as three quarters of the initial orbital frequency, which for the SXS data sets
is determined from the included metadata, and for the BAM data sets is computed
from the \PN initial data parameters for the \BH momenta. This value of
the  initial orbital frequency is cross-checked from the actual numerical
waveform data by extrapolating the frequency evolution of the numerical waveform
to the initial time by fitting to a fourth order polynomial in a time window
starting after a time $T_J$. As can be expected on dimensional grounds, this
procedure causes transients in the strain that persist for a time scale of the
order of $1/f_0$. We thus define
\begin{equation}
\Delta T_h = \sigma_h/f_0,   
\end{equation}
and discard numerical data before a time $t_0$,
where 
\begin{equation}
t_0= \max(\Delta T_h, \Delta T_J).   
\end{equation}
For short waveforms of less than $3000 M$ time to peak, we choose $\sigma_J= 1 , \sigma_h= 2.5$, and for longer waveforms we choose $\sigma_J= 3, \sigma_h=1.5$.

\subsection{Post-Newtonian and effective-one-body descriptions}
\label{sec:PN}

Our work uses \PN waveforms in two ways: First, to construct ``hybrid
waveforms'' that represent the complete inspiral-merger-ringdown of a binary
system by appropriately gluing together \PN and \NR waveforms. 
Without restricting the generality of our approach we will carry out this
procedure in the time domain (for an alternative frequency domain description
see \cite{Santamaria2010}). Second, we will tune a frequency-domain
\PN approximant to the hybrid waveforms by calibrating higher order
\PN terms (which have not yet been computed in \PN theory), in order
to represent the inspiral regime of our final IMR waveform model.

Post-Newtonian perturbation theory provides a rich zoo of different approximants, which solve the Einstein equations up to a certain order of expansion, but may differ in higher order terms depending on how intermediate non-linear mathematical operations are performed.
We will use standard quasi-circular non-precessing ``Taylor-approximants'',
described in \cite{Buonanno2009}, in particular the TaylorT1, TaylorT2 and
TaylorT4 time domain variants, and the TaylorF2 frequency domain version. These
are consistently derived from an energy and flux which incorporate up to 3.5 \PN
order non-spinning terms (see for instance \cite{Buonanno2009}), 3.5 \PN order
in spin-orbit interaction \cite{Bohe:2012mr,Bohe:2013cla}, 3PN order quadratic
in spin \cite{Bohe:2015ana}, and 3.5PN order cubic in spin \cite{Marsat2015}.

Terms corresponding to higher \PN order may also be added in order to conform
with a particular form of the energy, flux, or some other property of the
binary.
In our work we will in particular use models derived from the very successful
\EOB description \cite{Buonanno:1998gg,Buonanno:2000ef}.
In particular, we will use the SEOBNRv1 \cite{Taracchini:2012ig} and SEOBNRv2  \cite{Taracchini2014} models in two
forms, the proper SEOBNRv1/v2 waveform model, which is calibrated against numerical
relativity waveforms from the SXS catalogue \cite{SXS:catalog}, and an
uncalibrated version, which we refer to as SEOBv1/v2, where all terms calibrated to
\NR have been set to zero.
In detail, we have applied the following changes to the SEOBNRv2  model \cite{Taracchini2014}  to obtain the uncalibrated waveforms which we will refer to as SEOBv2: In the EOB Hamiltonian we have set the 4.5PN spin-orbit adjustable parametr $d_{SO}$  and 
the 3-PN spin-spin parameter $d_{SS}$ to zero, and we for non-spinning correction only keep the mass-ratio independent term in the parameter $K$, thus $K = 1.712$. The 3PN term added to the quantity $\delta_22$ in the factorized waveforms is also set to zero, as are all NQC corrections. Furthermore, in order to keep the implementation of the SEOBNRv2 in the file {\tt LALSimIMRSpinAlignedEOB.c} LAL library \cite{lalsuite} from crashing, the parameter {\tt tStepBack} was changed from {\tt tStepBack = 100.*mTScaled} to {\tt tStepBack = 300.*mTScaled}.
While this work has been in progress, it has been found in \cite{Szilagyi:2015rwa}, that an (independent) uncalibrated version of SEOBNRv2 shows excellent agreement with a 125-orbit-long mass-ratio 7 non-spinning waveform.

\subsection{Hybrids}\label{sec:hybrids}

We construct hybrid waveforms, where the early time behaviour is described by a
\PN approximant, and the late time behaviour by a \NR waveform, by essentially
standard procedures. See 
Refs.~\cite{Ajith:2007kx,Boyle:2011dy,MacDonald:2011ne,Ajith:2012az} 
for summaries of different methods in the literature, and 
Ref.~\cite{Bustillo:2015ova} for a detailed description of our specific procedure, which we briefly summarize here.

First, we note that any non-precessing-binary waveform $h(t)$ we consider represents an equivalence class of waveforms, which correspond to a binary with the same masses and spins, and only differ by a relative time shift, a relative rotation, and the sense of their orbital rotation (i.e. the sign of the frequency $\omega$). Without restricting generality we will choose the frequency as positive.
The \PN and \NR strains $h^{PN/NR}$ will then be related by
 \begin{equation}
h_{22}^{PN}(t)=e^{i\varphi_0}h_{22}^{NR}(t+\Delta t).
\end{equation}
In order to find good parameters $\Delta t$ and $\varphi_0$, we fix an initial time $t_0$ and fitting window length $T$, and minimize the quantity
\begin{equation}
\Delta(\Delta t; t_0,T)=\int_{t_0}^{t_0+T} \left( \omega^{PN}(t) - \omega^{NR}(t+\Delta t)\right)^2 dt,
\label{eq:omegaintegral}
\end{equation}
with respect to the time shift  $\Delta t$. Alternatively we replace $\omega$ by
$\phi$ to obtain a consistency and robustness check, which is particularly
useful when automatically processing tens of waveforms as we do. Our original
choice of $\omega$ has the  advantage of not depending on $\varphi_0$. Other
authors have replaced $\omega$ by $h$ \cite{MacDonald:2011ne} in the integrand,
here we only use phase and no amplitude information in order to avoid the
influence of amplitude errors. In \cite{Bustillo:2015ova} we have checked that
the window length $T$ that corresponds to four \GW cycles (two orbits) is
sufficient to average out oscillatory waveform artefacts mostly due to residual
eccentricity. In order to find an appropriate phase shift $\varphi_0$ for given
$\Delta t$ from the minimization of 
Eq.~(\ref{eq:omegaintegral}), we align the phases at the middle of the matching window.

The hybrid waveform is then defined as,
\begin{equation}
h(t) = \omega^-_{(t_0,t_0+T)}  e^{i\varphi_0}h^{PN}(t+\delta t) + \omega^+_{(t_0,t_0 + T)}  h^{NR}(t),
\label{eq:hybrid22}
\end{equation} where $\omega^{-/+}$ are linear transition functions over the time window $t \in [t_0, t_0 + T]$.

The crucial step, however, when  ``glueing'' \PN to numerical waveforms is to
consistently align them at some reference frequency, where we choose their time
coordinates and phases to coincide. If the waveforms are identical, the relative
time (and phase) shift between them will be independent of the matching
frequency. We are interested in the case where the waveforms are different, and
where we have the choice of \PN approximant, e.g.~one discussed in
Sec.~\ref{sec:PN}, to attach to a particular numerical waveform.  For
non-identical waveforms the relative time-shift between the \PN and NR
descriptions will depend on the frequency where the two waveforms are compared
or matched together. A preferred \PN approximant would exhibit a time shift
relative to the \NR waveform which depends only negligibly on time.
In order to find the best-matching \PN approximant, we compute the hybridisation
time shift as a function of matching frequency as follows. We consider the time
dependent frequencies $\omega_{PN}(t)$ and $\omega_{NR}(t)$. We can invert these
functions numerically, e.g. by evaluating them for a series of time steps, and
then interpolating inverse functions of $t(\omega)$. To compute the time shift
we then compute the difference 
\begin{equation}\label{eq:DeltaT}
\Delta t =  t_{PN}(\omega)-t_{NR}(\omega).
\end{equation}
The results are displayed in Figs.~\ref{fig:DT_q1NS}-\ref{fig:DT_q1_q18}.
First, in Fig.~\ref{fig:DT_q1NS} we consider the equal mass non-spinning case,
where we however use SEOBNRv2 and not the \NR results as the reference waveform,
in order to reduce the effect of oscillations in the \NR result due to residual
eccentricity. One can see that the \NR waveform 
SXS:BBH:0001 
from the SXS catalogue, the uncalibrated SEOBv2 and TaylorT4 are very close to
each other over the frequency range of the \NR waveform. SEOBNRv2 and SEOBv2 are
close over the whole frequency range plotted, $0.01 \leq M \omega_{22}$.

In the left panel of Fig.~\ref{fig:DT_q1_q18} we consider the BAM mass-ratio 18 case with a mild dimensionless spin of $\chi_1 = 0.4$ parallel to the orbital angular momentum on the bigger BH, and a non-spinning smaller hole. Again we use SEOBNRv2 as the reference waveform.
We see that the \NR and SEOBNRv2 waveforms agree well over the common frequency
range, and the approximant that performs next best is the uncalibrated SEOBv2. Note that now the $\Delta t$ time scales are much larger than in Fig.~\ref{fig:DT_q1NS}, and that the standard \PN approximants perform markedly
worse in comparison to the \NR than the SEOB(NR)v2 waveforms. 

\begin{figure}[htbp]
\includegraphics[width=\columnwidth]{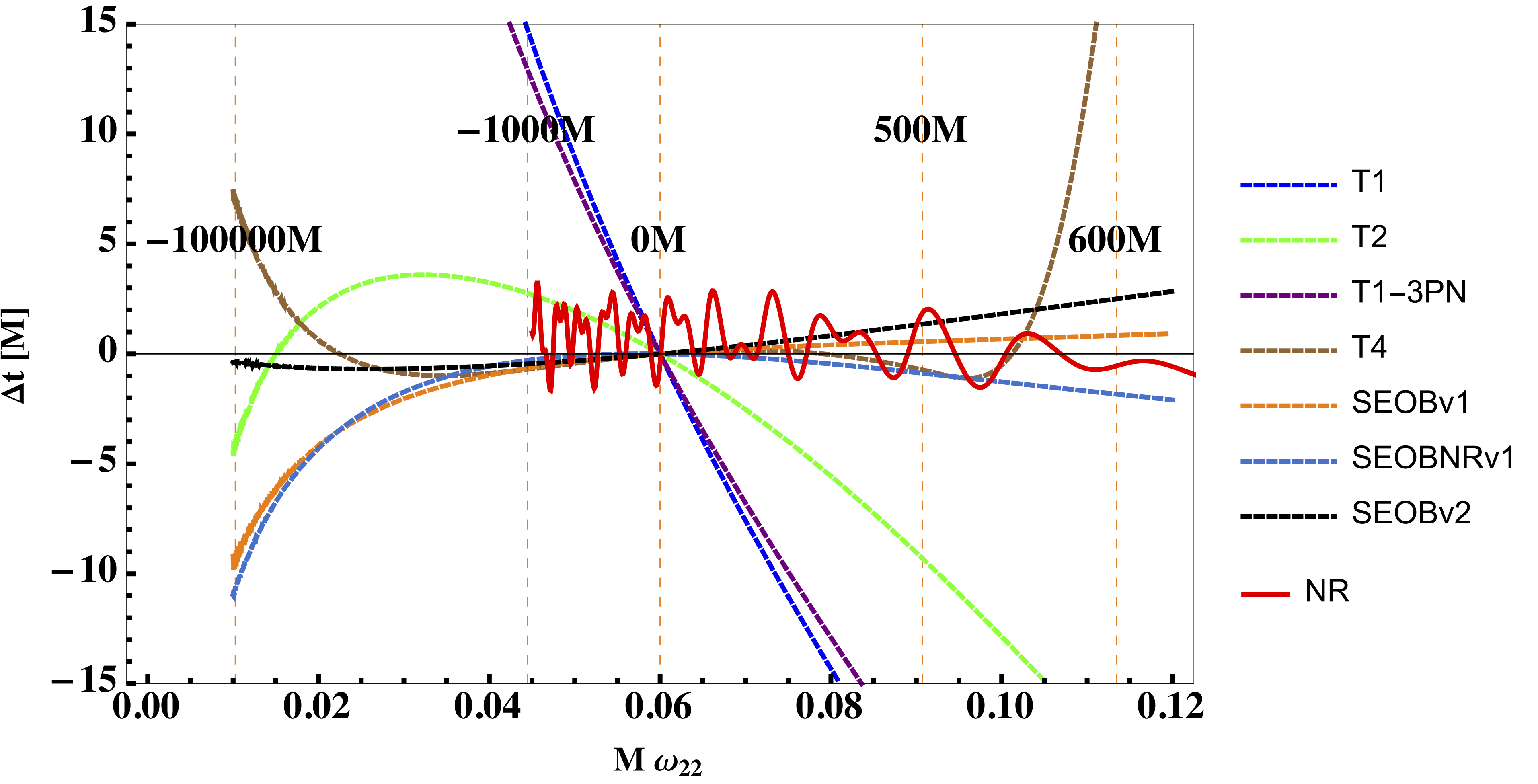}
\caption{Time shift $\Delta t$ according to Eq.~(\ref{eq:DeltaT}) as a function
of wave frequency $M\omega$ for the case of non-spinning $q=1$, using SEOBNRv2
as the reference waveform and SXS:BBH:0001 from the SXS catalogue as the \NR
waveform. As is well known, the SEOBNRv2 model and TaylorT4 are very close to
the \NR result in this case. Uncalibrated SEOBv2 (uncalibrated SEOBNRv2) is also
very close, while TaylorT1, TaylorT1 cut off at 3PN order, and TaylorT2 show
significant dephasing. The SEOBNRv1 and SEOBv1 (uncalibrated SEOBNRv1) show rather
similar performance, with good agreement with NR, and disagreement with SEOBNRv2 at lower frequencies.
}
\label{fig:DT_q1NS}
\end{figure}

\begin{figure*}[htbp]
\includegraphics[width=\columnwidth]{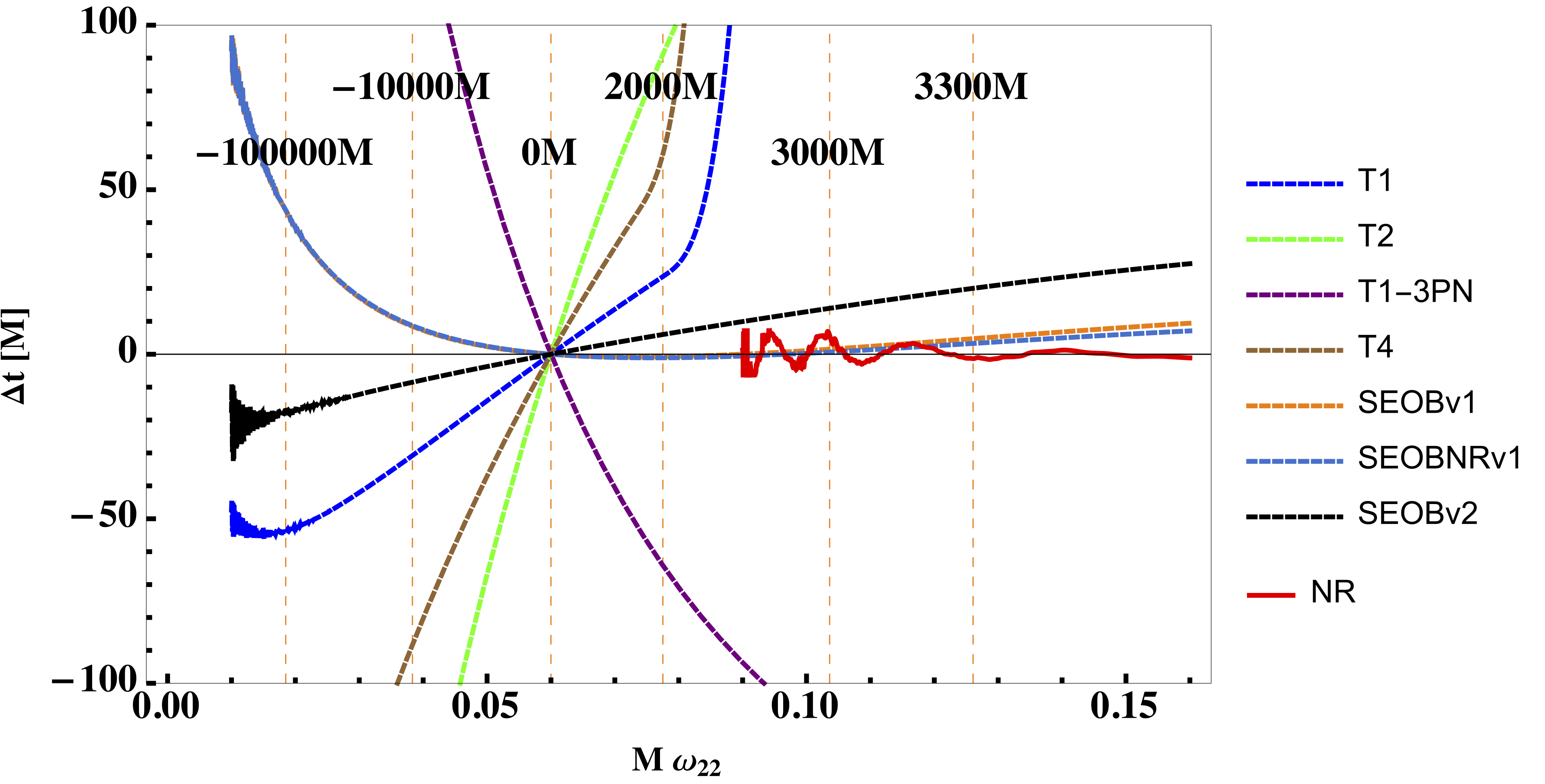}
\includegraphics[width=\columnwidth]{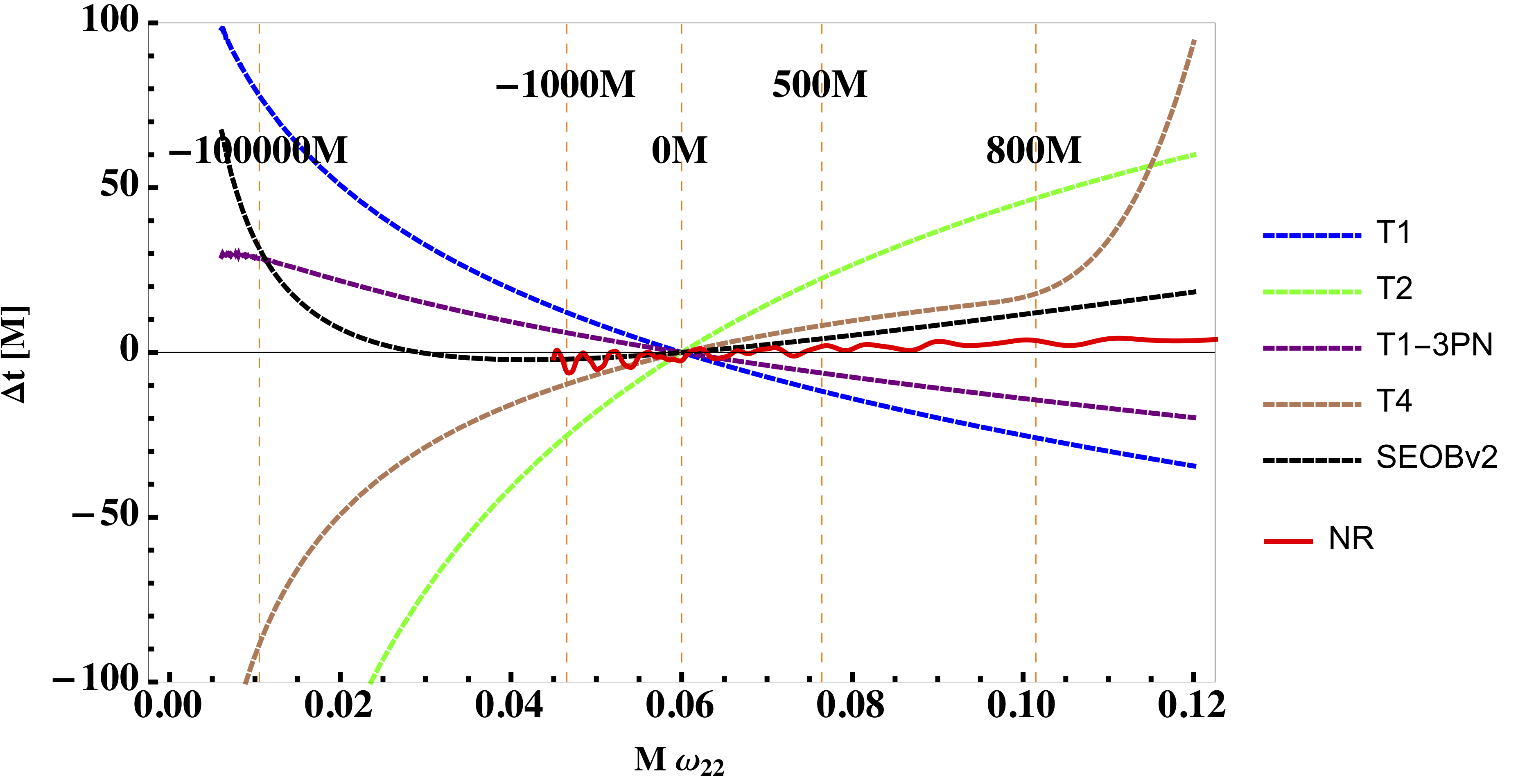}
\caption{Left panel: Time shift $\Delta t$ according to Eq.\ref{eq:DeltaT} as a function of
wave frequency $M\omega$ for the case of mass ratio 18 with spins $\{0.4, 0\}$ with
SEOBNRv2 as the reference waveform. The PN/EOB waveforms start at $M \omega=0.01$. Right panel:
Same analysis for an equal mass binary with equal spins $\chi=0.98$. Tthe \NR waveform
is SXS:BBH:0172 from the SXS catalogue, see Table \ref{tab:SXSwftable}. SEOBNRv1/SEOBv1 are not shown for this case, since
the model is not valid for high spins parallel to the orbital angular momentum.}
\label{fig:DT_q1_q18}
\end{figure*}

In the right panel of Fig.~\ref{fig:DT_q1_q18} we finally consider again equal masses, but now
with large spins of $\chi_1 = \chi_2 = 0.98$ parallel to the orbital angular
momentum. The \NR waveform is 
SXS:BBH:0172, produced with the SpEC code. The dephasing between the \NR
waveform and SEOBNRv2 is again rather small, although significantly larger than
in the nonspinning case. The SEOBv2 waveform is the approximant closest to
SEOBNRv2 for waveforms not longer than $10^5 M$. For lower frequencies, TaylorT1
cut off at 3PN order is closer to SEOBNRv2, but all other approximants show
significantly larger differences. 

The time-shift analysis discussed here for three examples has also been carried
out for a number of other cases, with similar results, which will be presented
in more detail in a separate paper. The conclusion for our current work is that
in the frequency regime relevant for hybridisation, SEOBv2 uncalibrated \EOB 
waveforms are almost as close to \NR results as the calibrated \EOB  model, and
at low frequencies are typically closer to SEOBNRv2 than any other approximant
at 3.5PN order. In Ref.~\cite{Bustillo:2015ova} we
have shown a similar result, where we have applied a similar analysis to compare
TaylorT1, TaylorT4 and SEOBNRv1 to the BAM non-spinning \NR waveform at mass
ratio 18, and have found that SEOBNRv1 is far superior for hybridization.

In this work we will therefore construct all our hybrid waveforms, which form the input to our phenomenological modeling, with the SEOBv2 model. We are thus able to calibrate an inspiral model that is independent from the calibration process performed for SEOBNRv2, which will allow valuable cross checks.

Frequency domain strain hybrids are constructed by sampling the time domain
hybrid with equispaced time steps, and aligning all waveforms such that the
value of the time coordinate and phase at the peak amplitude vanish. The
waveforms are then zero-padded to all start at a time $t_{\rm start}$ and all
end at a time $t_{\rm end}$, where $t_{\rm start}$ is significantly smaller than
the beginning of the numerical \EOB  data, and $t_{\rm end}$ is chosen at
several hundred $M$ after the amplitude peak. Different batches of hybrids have
been produced for comparisons, with the time-step typically chosen 
as $0.5 M$ or $1M$. 

The time-domain amplitude for selected hybrids is shown in Fig.~\ref{fig:TDAmplitude_cornercases}. The strain for
the same cases is shown in Fig.~\ref{fig:4WFs}, and the Fourier-domain amplitudes are shown in Figs.~\ref{fig:Amp_Corner}.

\begin{figure}[htbp]
\includegraphics[width=\columnwidth]{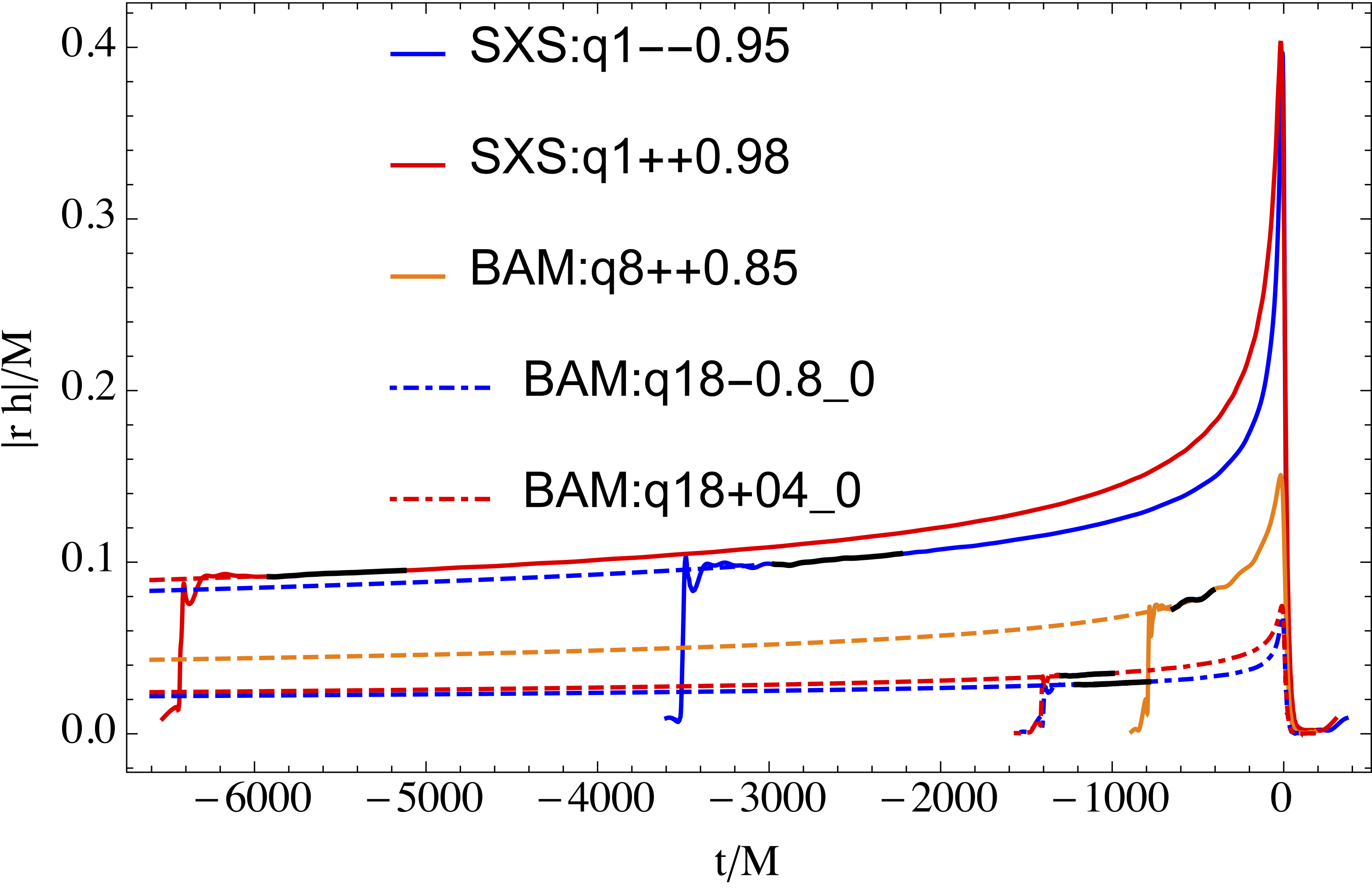} 
\caption{
Time domain amplitude of configurations at the corners of our parameter-space
coverage, aligned at merger, showing both the \NR waveform and the hybrid
(marked as dashed lines). Thick black lines mark the hybridization regions.}
\label{fig:TDAmplitude_cornercases}
\end{figure}

\begin{figure*}[htbp]
\includegraphics[width=2\columnwidth]{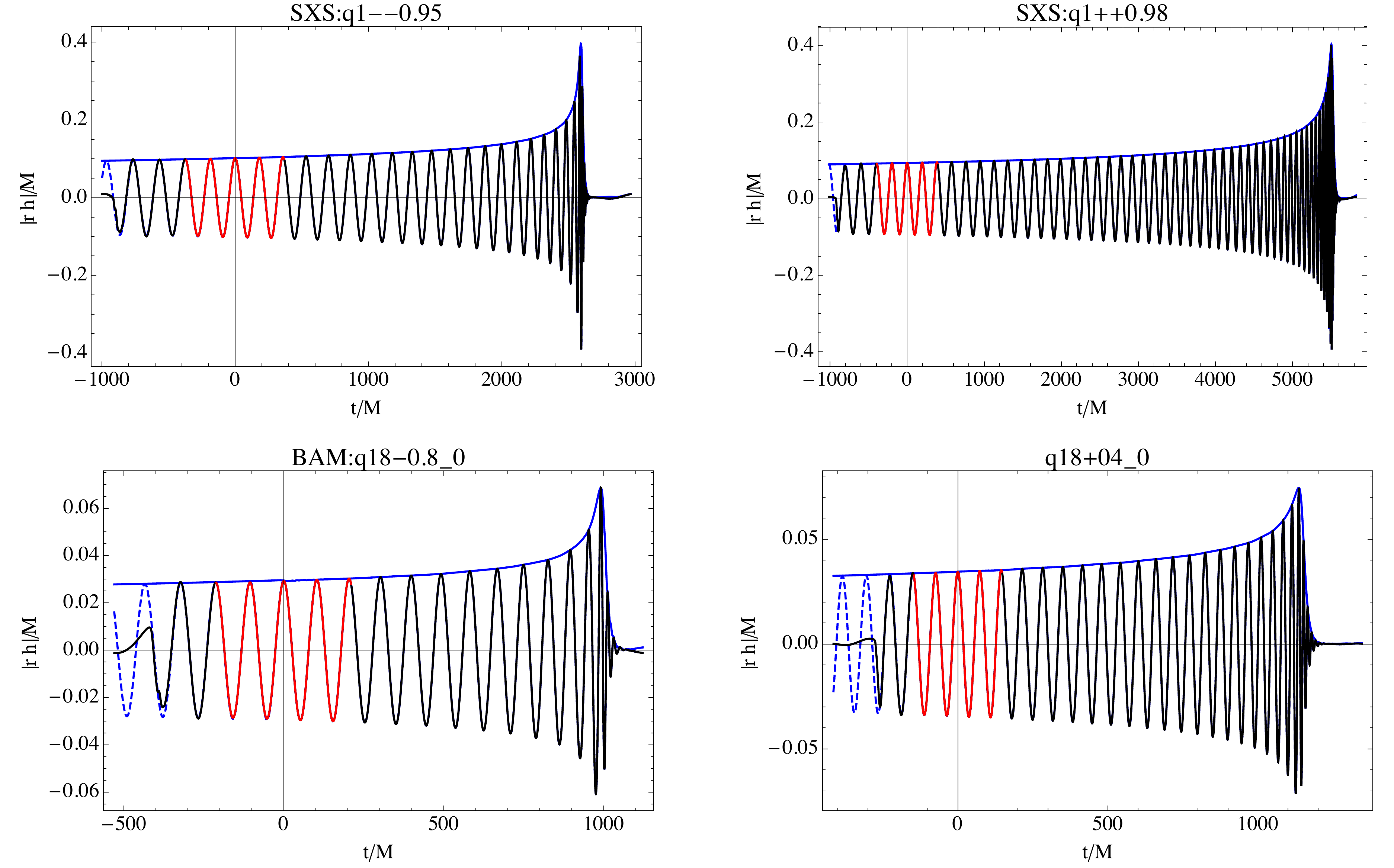} 
\caption{
Real part of the time domain strain of corner cases, time coordinate is set to zero at the mid-frequency of the hybridization region, which is shown in red.}
\label{fig:4WFs}
\end{figure*}

\begin{figure*}[htbp]
\includegraphics[width=\columnwidth]{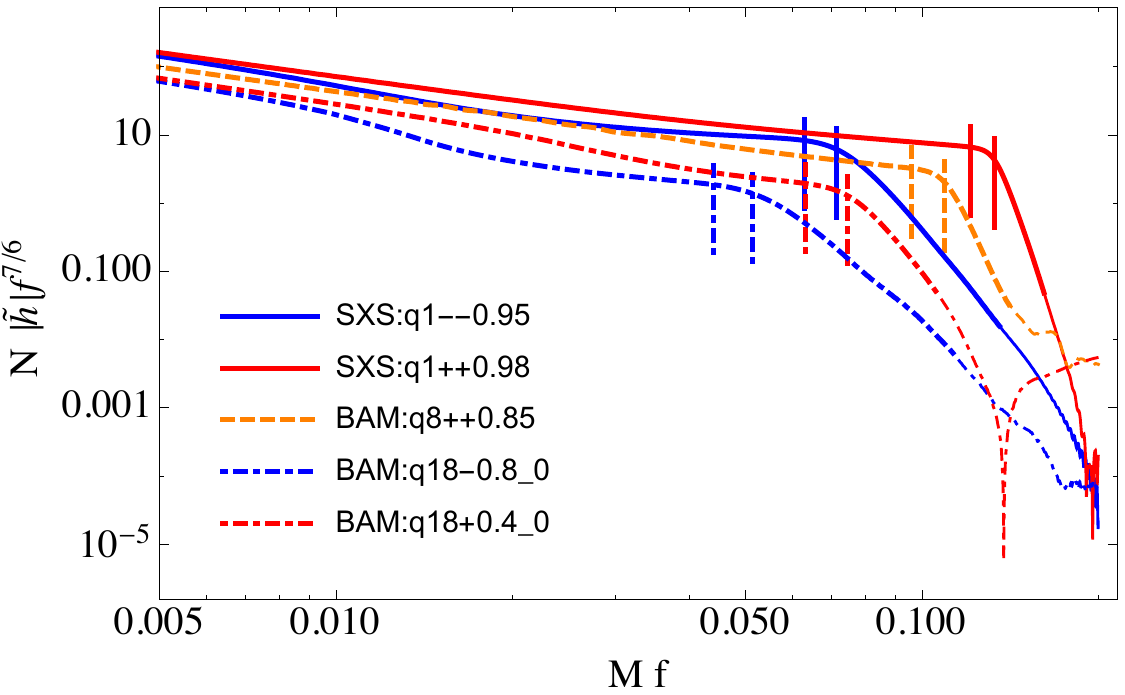}       
\includegraphics[width=\columnwidth]{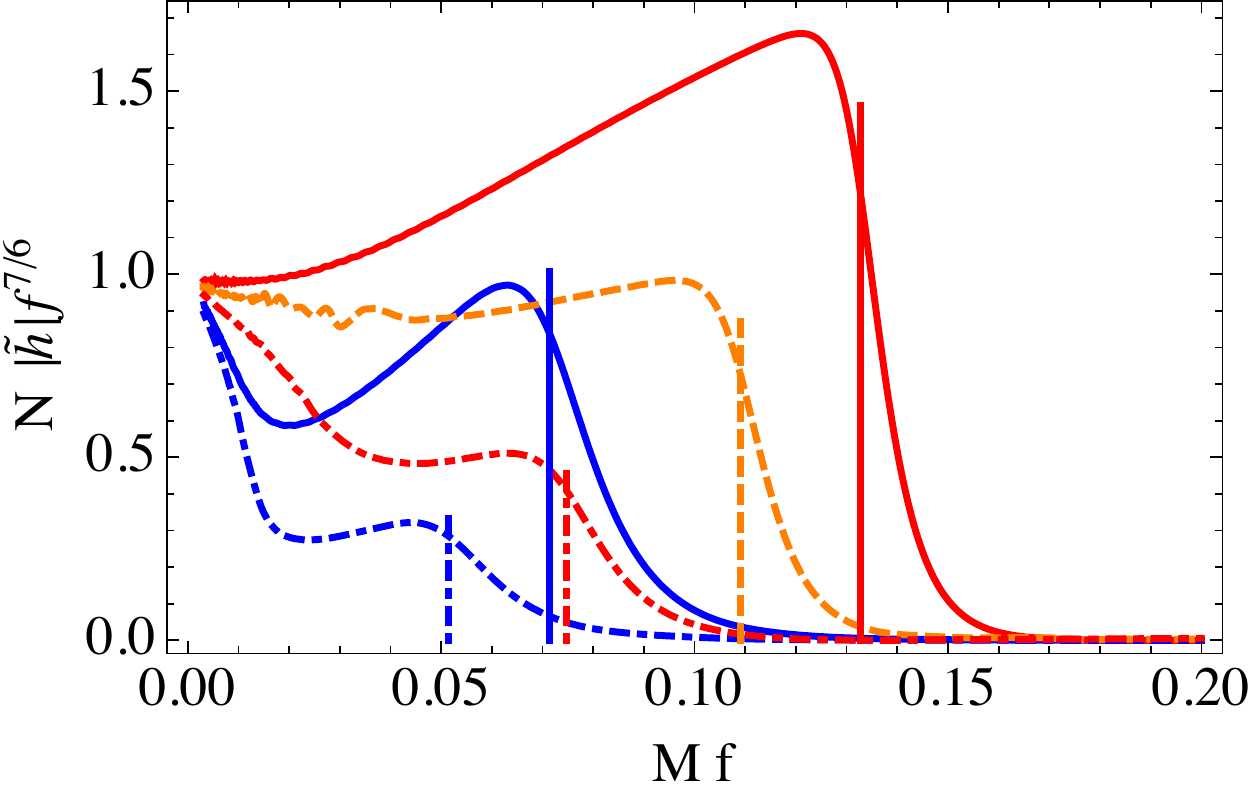}  
\caption{Fourier domain amplitude of corner cases and ringdown frequency. In the
left panel, the thick lines indicate the part of the \NR waveform that was used
in the final hybrid, while data indicated by a thin line contained noise or
other spurious artifacts,
and was discarded. The vertical lines indicate the peak of the amplitude, and the ringdown frequency $f_{RD}$. 
In the right panel the $f^{-7/6}$ leading-PN-order amplitude has been scaled out to
illustrate detailed features of the Fourier-domain amplitude through merger and ringdown.
}
\label{fig:Amp_Corner}
\end{figure*}

\section{Modeling the final state}
\label{sec:finalstate}

While it has not been rigorously proven within general relativity, the final
state of the coalescence observed in all \NR calculations is a single perturbed
Kerr \BH. The complex frequency of the ``quasinormal'' ringdown provides two key
pieces of information for modelling the \GW signal, and can be computed directly
as a function of the final mass and spin. As a first step to model waveforms, we
will thus model the final state and provide simple analytical fitting formulas
for the radiated energy and final spin as a function of symmetric mass ratio and
an effective spin parameter. These cover a wider parameter space than previous work in the literature;
see Refs.~\cite{Barausse:2009uz,Hemberger:2013hsa,Healy:2014yta,Zlochower:2015wga} and references therein.
This
will also illustrate how we construct analytical fits across the parameter space, and similar procedures will be used afterwards in our waveform modelling. The main goal of our approach to model the final state is to construct accurate explicit expressions, which are therefore fast to evaluate and easy to generalize to larger calibration data sets.

To determine the dimensionless ringdown frequency $M \omega_{RD}$ as a function of $a_f$, we have interpolated the data set in Refs.~\cite{Berti:2009kk,BertiWebData}. These well known functions are shown in Fig.~\ref{fig:RDfreqs}, together with the wave frequency corresponding to a test particle on innermost stable orbit (ISCO) \cite{Bardeen:1972fi} of Kerr spacetime. One can see that the steepest functional dependence happens for large positive spins.

To calculate the dimensionful ringdown frequency of the binary, loss of energy during inspiral needs to be accounted for as
 \begin{equation}
\omega_{RD} = \frac{ (M\omega_{RD})}{M_f} = \frac{ (M\omega_{RD})}{M - E_{\mbox{rad}}},       
 \end{equation}
and we need to model the radiated energy $E_{\mbox{rad}}$ in addition to the final Kerr parameter.
 
\begin{figure}[htbp]
\includegraphics[width=\columnwidth]{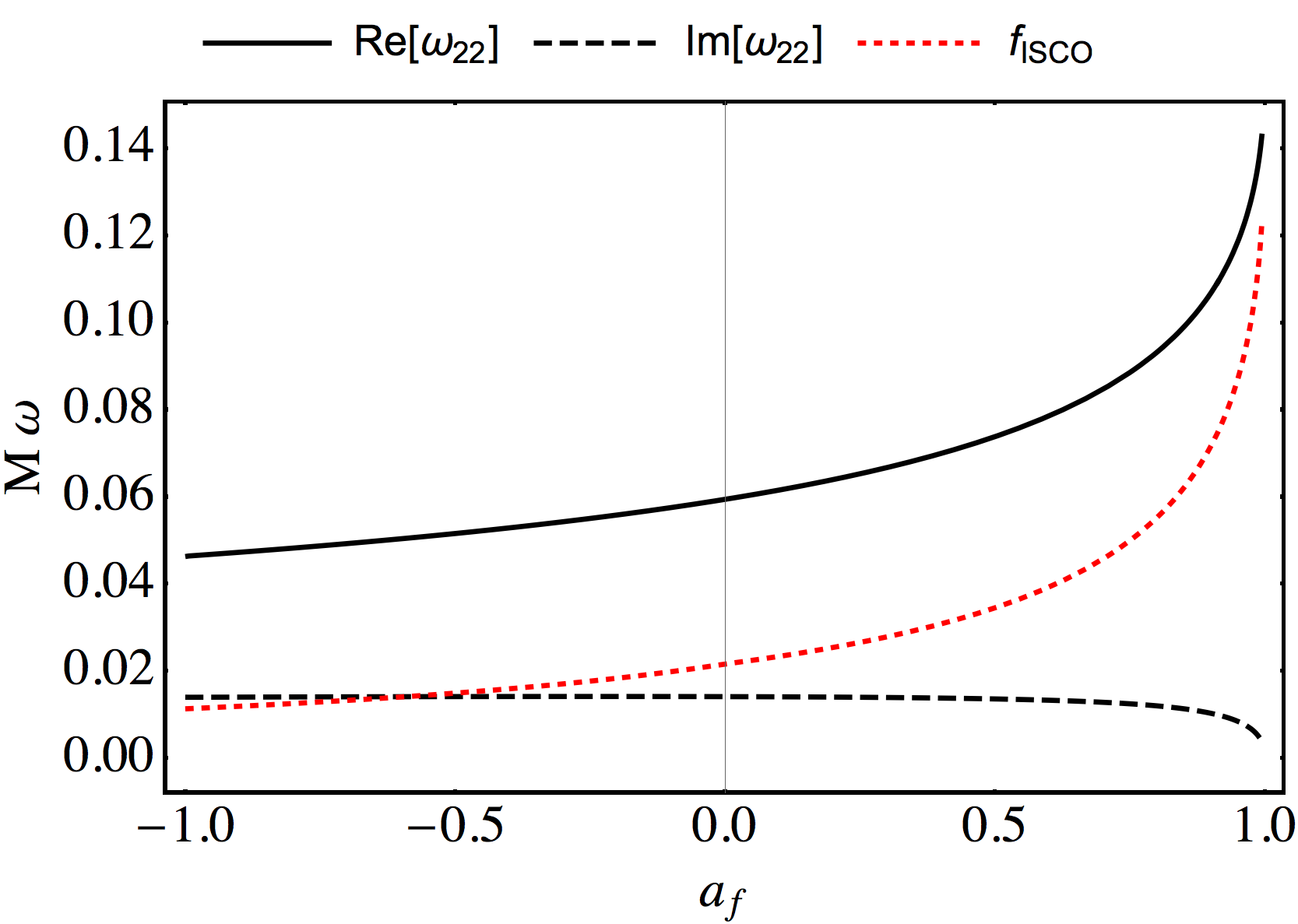}
\caption{Real (oscillatory)and imaginary (damping) part of the ringdown
frequency as a function of the final Kerr parameter $a_f$. A positive sign
denotes a final \BH parallel to the initial orbital angular momentum, and
negative sign antiparallel. 
}
\label{fig:RDfreqs}
\end{figure}

We have used final spin and radiated energy results from the following 106 data sets:
15 new BAM cases at mass ratios 2, 3, 4, 8, 18,
5 old BAM cases at mass ratio 4 from published in \cite{Puerrer2013} (used only for the final spin),
50 cases from the SXS catalogue ~\cite{SXS:catalog}, and 
36 cases from the RIT group \cite{Healy:2014yta}. For the RIT runs, two type of results are given, derived from the isolated horizon formalism and from the radiation, and we use the isolated horizon quantities due to the lower error bars quoted. We do not use the mass-ratio 10 results from the RIT data set, since the quoted result for the radiated energy differs from ours and the numerical fit by a factor of two and is possibly a misprint.

\subsection{Final spin}
\label{sec:finalspin}
 
Before merger, the total angular momentum can be expressed in terms of the individual BH spins $\vec S_{i}$, which for our purposes change only negligibly during inspiral, and the orbital angular momentum $\vec L$ as,
 \begin{equation}\label{eq:JLSS}
 \vec J = \vec L + \vec S_1 + \vec S_2.
\end{equation}
Due to symmetry all angular momentum vectors are orthogonal to the orbital plane and the only non-zero component is the $z$-component. For simplicity of notation, 
we therefore drop the vector notation, and write $J, S_i$, etc., for the  $z$-component.
We seek to approximate $J$, $L$, and correspondingly the final Kerr parameter $a_f = J_f/M_f^2$ through a function of the mass ratio and a 
single effective spin parameter.
Given that Eq.~(\ref{eq:JLSS}) depends trivially on the sum of the individual spins, we attempt the approximation
$$
 \vec J_f \approx \vec J_f(\eta, \vec S), \quad  \vec S = \vec S_1 + \vec S_2.
$$
In the infinite mass ratio limit, $\eta=0$, the final spin coincides with the spin of the larger BH,
\begin{equation}
J_f(\eta, S_1, S_2) = S_1 = S, \quad  a_f(\eta, \chi_1, \chi_2) = \chi_1.
\end{equation}
We will frequently use a rescaled version of our total spin $S = S_1 + S_2$ to $\hat S = S/(1-2\eta)$, for which $-1 \leq \hat S \leq 1$, consistent with the extreme Kerr limit.

The data points are shown in Fig.~\ref{fig:afinal}. As can be seen, our waveform coverage is densest at equal mass and sparser at higher mass ratios. 
To produce an analytical fit, we first inspect the data set displayed in Fig.~\ref{fig:afinal}, and in particular the non-spinning and equal mass subsets. We find that
fourth order polynomials in $\eta$ and $S$ produce accurate fits for the two subsets, and we fix the linear term in $\eta$ by a Taylor expansion around the extreme mass ratio limit as in \cite{Rezzolla:2007rd},
\begin{equation}
a_f = 2 \sqrt{3} \eta + \mbox {higher order in $\eta$ and spins}.
\end{equation}
In order to cover the whole parameter space we extend the ansatz by terms quadratic in $\eta$ and quartic in the total spin $S$ to 
\begin{eqnarray}
a_f ^{eff} &=& f_ {00}  + S + \eta \sum_ {i = 0, 4}\frac {f_ {1 i}} {i!} S^ i \nonumber\\
& &+ \eta^2  \sum_ {i = 0, 4}\frac {f_ {2 i}} {i!} S^i +  f_ {30}\eta^3 + f_ {40}\eta^4.
\end{eqnarray}
After fixing coefficients for consistency with the nonspinning and equal mass cases, the only coefficients left to fit are the 4 numbers
$(f_{11}, f_{12}, f_{13}, f_{14})$.
The result is,
\begin{eqnarray}
&a_f ^{eff}(\eta, S)  =  
S  + 2 \sqrt3 \, \eta - 4.399 \, \eta^2 + 9.397 \, \eta^3 - 13.181 \, \eta^4 & \nonumber \\
& + (-0.085 \, S + 0.102 \, S^2 - 1.355 \, S^3 - 0.868 \, S^4) \, \eta \nonumber \\
& + (-5.837 \, S - 2.097 \, S^2 + 4.109 \, S^3 + 2.064 \, S^4) \, \eta^2
\label{eqn:FinalSpin}
\end{eqnarray} 
RMS errors are $6.8 \times 10^{-3}$, and $2.4 \times 10^{-3}$ when comparing the fit with the equal spin subset. We also 
note that this fit respects the Kerr limit, i.e., $a_f^{\rm eff}| \leq 1$.

\begin{figure}[htbp]
\includegraphics[width=\columnwidth]{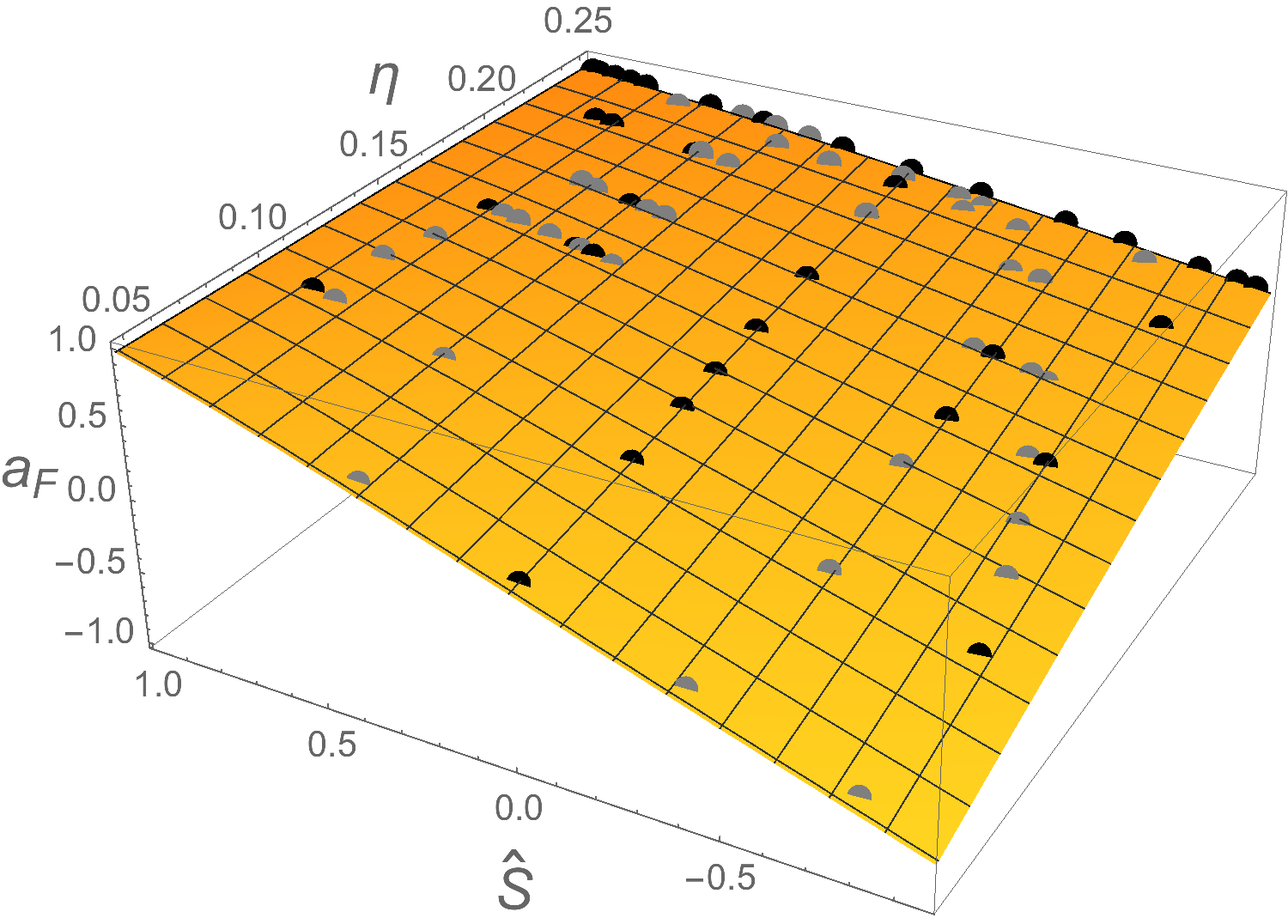}
\caption{Final Kerr parameter from Eq.~\ref{eqn:FinalSpin} plotted as a function of symmetric mass ratio and total spin $\hat S$. Black dots mark data points with equal spins, grey dots unequal spins.}
\label{fig:afinal}
\end{figure}

\subsection{Radiated energy}
\label{sec:erad}

The data points for radiated energy are shown in Fig.~\ref{fig:afinal_extremeKerr}, 
together with the effective single spin fit we will now discuss. Inspecting the plot, clearly, a polynomial in the symmetric mass ratio and the effective spin will not provide the ideal model, and a rational function model comes to mind. Also, as in the final spin case, clearly a reliable accurate model of the whole parameter space would require further data points for large spins.

It turns out that after factoring out a fit to the nonspinning subset, the radiated energy depends only rather weakly on the symmetric mass ratio. We find that the non-spinning radiated energy, $E_{rad}^{NS}(\eta)$, is very well captured by a fourth order polynomial, at a RMS of $3.5 \times 10^{-5}$,
\begin{equation}
E_{rad}^{NS}(\eta) = 0.0559745 \eta + 0.580951 \eta^2  - 0.960673 \eta^3 +  3.35241 \eta^4.
\end{equation}

As can be seen in Fig.~\ref{fig:afinal}, the effective spin $\hat S$ works reasonably well for the radiated energy. We model the dependence on the spin $\hat S$ through
a simple rational function, where the numerator and denominator are linear in spin and quadratic in symmetric mass ratio, with some experimentation having gone into making a choice for which the nonlinear fitting procedure converges well, and the result has no singularities due to vanishing denominator.
The result of the fit, with a RMS error of $4 \times 10^{-4}$ is,
\begin{equation}
\frac{E_{rad}}{M_{ini}} =  E_{rad}^{NS}(\eta)  
\frac{1+ \hat S \left(-0.00303023 - 2.00661 \eta + 7.70506 \eta^2 \right)}{1+\hat S \left(-0.67144 - 1.47569 \eta + 7.30468 \eta^2 \right)}
\label{eqn:Erad}
\end{equation}

\begin{figure}[htbp]
\includegraphics[width=\columnwidth]{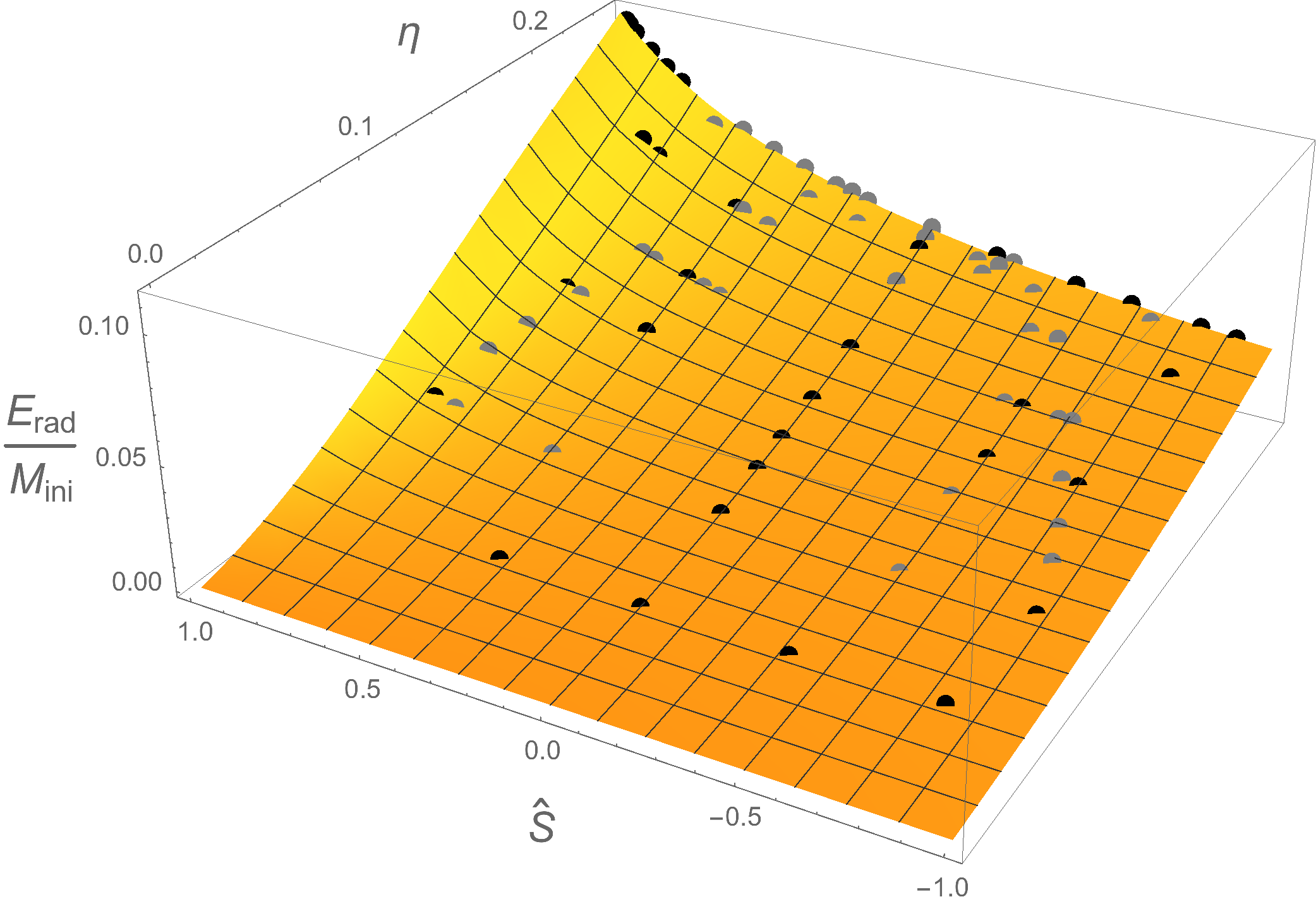}
\caption{Radiated energy according to fit Eq.~\ref{eqn:Erad} plotted as a function of symmetric mass ratio and total spin $\hat S$. Black dots mark data points with equal spins, grey dots unequal spins.}
\label{fig:afinal_extremeKerr}
\end{figure}

\section{Waveform anatomy and model} 
\label{sec:model}

\subsection{Waveform anatomy}

\subsubsection{Amplitude}\label{sec:amp}

The waveform anatomy in the time domain has been studied extensively, and
combined with \EOB  resummation techniques of \PN results has given rise to the
family of \EOB-\NR waveform models
\cite{Buonanno:2007pf,Damour:2008te,Pan:2011gk,Damour:2012ky,Taracchini2014,
Nagar:2015xqa}. One may distinguish three phases without sharp boundaries: (i) a
long inspiral with slowly increasing amplitude, where the amplitude scales as $M
\eta \omega^{2/3}$ in the low frequency limit, and the coalescence time as
$\omega^{-8/3}/\eta$, (ii) an ``extended-merger'' characterised by a rapid
increase in amplitude and frequency; (iii) followed by a damped sinusoidal
ringdown. The Fourier transformation to obtain the frequency domain waveform can
in general not be carried out analytically.  In the low-frequency regime
however, the stationary phase approximation (SPA) can be used to analytically
obtain an approximate Fourier transform, which is used in particular to obtain
the TaylorF2 \PN approximant as a closed-form expression. 
Following the procedure outlined for example in Section 3 of Ref.~\cite{Santamaria2010} we
obtain the SPA amplitude using the TaylorT4 form of the energy balance equation.
The Fourier domain amplitude, $\tilde{A}_{22}$, is then given in terms of the
time domain amplitude $A_{22}$ and second phase derivative $\ddot{\phi}$
evaluated at the Fourier variable $t_f = (2 \pi f/m)^{2/3}$,
where $m=2$ for the dominant harmonic we consider,
\begin{equation}
\tilde{A}_{22}(f) = A_{22}(t_f) \sqrt{\frac{2 \pi}{m \ddot{\phi}(t_f)}} \, .
\label{equ:xT4Amp}
\end{equation}
In particular, to leading order the Fourier amplitude is,
\begin{equation}\label{eq:T2LeadingOrder}
\vert \tilde h_{22} \vert = A_0 f^{-7/6} (1 + O(f^{2/3})), \qquad  A_0 = \sqrt{\frac{2 \, \eta}{3 \, \pi^{1/3}}}.
\end{equation}

In order to better emphasize the non-trivial features of the amplitude, we rescale our numerical data sets by the factor ${f^{7/6}}/{A_0}$,
to normalize all amplitudes to unity at zero frequency as shown in Fig.~\ref{fig:Amp_Corner}. 
We see a structure that is sufficiently rich that a single analytical expression
for the entire frequency range is difficult to achieve in terms of elementary
(and thus computationally cheap) functions. Our strategy will thus split our
description into an inspiral part, which models the waveform as higher order
corrections to \PN expressions, a merger-ringdown part which builds upon the
knowledge of the final state, and an intermediate part which describes the
frequency regime which can not be based directly upon \PN or the final state.

Regarding the merger-ringdown, a crude time domain model, which can be Fourier-transformed analytically, is a sine-Exponential, which is symmetric around the peak amplitude:
\begin{equation}
h(t) = e^{2 \pi (i f_{\text{RD}} t - f_{\text{damp}} \vert t \vert)}  \qquad f_{\text{RD}}, \, f_{\text damp} \in R.
\end{equation}
The Fourier transform (\ref{eq:def:Fourier}) yields a Lorentzian,
\begin{equation}\label{eq:defLorentzian}
\tilde h(\omega) = -\frac{1}{\pi}\frac{f_{\text{damp}}}{(f -f_{RD})^2+f_{\text{damp}}^2},
\end{equation}
which only falls off as $f^{-2}$ at large frequencies as expected from Eq.~(\ref{eq:FourierFalloff}), due to the fact that the original time domain waveform $h(t)$ is only $C^0$ at the peak. Despite its oversimplification, in particular the unphysical symmetry around the peak, and the incorrectly slow falloff at high frequency, the Lorentzian has provided a valid model for frequencies higher than the ringdown frequency in PhenomA/B/C models. Looking at Fig.~\ref{fig:Amp_Corner}, the expected roughly exponential drop at high frequencies are clearly visible.

A natural extension of the Lorentzian ansatz used in the previous Phenom models, is to model the merger-ringdown amplitude $A_{MR}$ by multiplying the Lorentzian by an exponential as
\begin{equation}\label{eq:AmpAMR}
\frac{A_{\text{MR}}}{A_0} = \gamma_{1} \frac{(\gamma_{3} f_{\text{damp}})}
                {(f-f_\text{RD})^2+(f_{\text{damp}}\gamma_{3})^2}
                e^{-\lambda(f-f_{\text{RD}})} \, .
\end{equation}
In order to find best fit parameters, we use Mathematica's {\tt NonlinearModelFit} function. To achieve robust convergence of the nonlinear least squares fit, we redefine 
$$
\lambda = (\gamma_{2}/(f_{\text{damp}}\gamma_{3})),
$$
to achieve a magnitude of order unity for the new free parameter $\gamma_2$. The merger-ringdown amplitude model thus has three free parameters that are fit from numerical waveform data, and the real and imaginary part  $f_{\text{RD}}, f_{\text{damp}}$ of the QNM ringdown frequency are computed from the known fits for final spin and mass from 
Eqs.~(\ref{eqn:FinalSpin}) and (\ref{eqn:Erad}) and the known dependence of the QNM frequencies on the dimensionless Kerr parameter shown in Fig.\ref{fig:RDfreqs}.  Interestingly, it turns out that $\gamma_3 \in [1.25,1.36]$, i.e., it has a very narrow dynamical range and could also be approximated as constant $\gamma_3  \approx 1.3$.

The exponential factor in our ansatz  Eq.~(\ref{eq:AmpAMR}) shifts the amplitude peak from  $f_{\rm RD}$ for the Lorentzian to
\begin{equation}
f_{\rm max} = \left|  f_{\text{RD}} + \frac{f_{\text{damp}} \gamma_3 \left( \sqrt{1 - \gamma_2^2} -1  \right)}{\gamma_2}  \right|.
\end{equation}

An appropriate frequency range for fitting the model Eq.~(\ref{eq:AmpAMR}) to numerical data can be determined as follows. At the large frequency end, we first discard frequencies with significant numerical noise by the following heuristic procedure: We multiply the amplitude with $f^p$, where we find that $p \approx 4$ works well. For physically accurate data we expect a monotonic decrease for the rescaled amplitude. Due to unphysical artefacts however, a minimum will appear, and we discard all frequencies higher than the one of the minimum. The result can be seen in Fig.~\ref{fig:Amp_Corner}. 
While no ``fine-tuning'' is required, a good choice of frequency range for model
fitting will not only increase the accuracy of the fit, but also lead to
smoother variation over the \BH parameter space of mass ratio and spins, which
will be essential when constructing a phenomenological model in Paper 2
\cite{Khan2015}. At low frequency we find that the fit is only good up to the
peak $f=f_{\rm max}$, while at large frequencies, where the amplitude is already
very small, we need to control contamination from numerical artefacts. Also, to
avoid boundary effects, we extend the fitting range to slightly smaller
frequencies than $f=f_{\rm max}$.  We find that two strategies work well:
specifying the fitting range in terms of multiples of the ringdown frequency and
the numerically determined local maximum, e.g.~using the range $(f_{\rm max},1.2
f_{\rm RD})$, or using the width of the Lorentzian, e.g.~as $(f_{\rm max} -
f_{\rm damp}/2 ,f_{\rm RD} + 2  f_{\rm damp})$. Fits and residuals for the
latter choice are shown in Figs.~\ref{fig:RDAmpComparison} and
\ref{fig:RDAmpResiduals}.

Since the merger-ringdown amplitude model Eq.~(\ref{eq:AmpAMR})  thus covers a
frequency range which will be covered even by the shortest \NR simulations of
\BH mergers, simulations only need to be sufficiently long to control the
systematic error of residual eccentricity, to avoid low frequency artefacts from
the Fourier transformation, and to clearly separate the unphysical radiation
content discussed in Sec.~\ref{sec:NR} from the physical part of the signal.

\begin{figure}[htbp]
\includegraphics[width=\columnwidth]{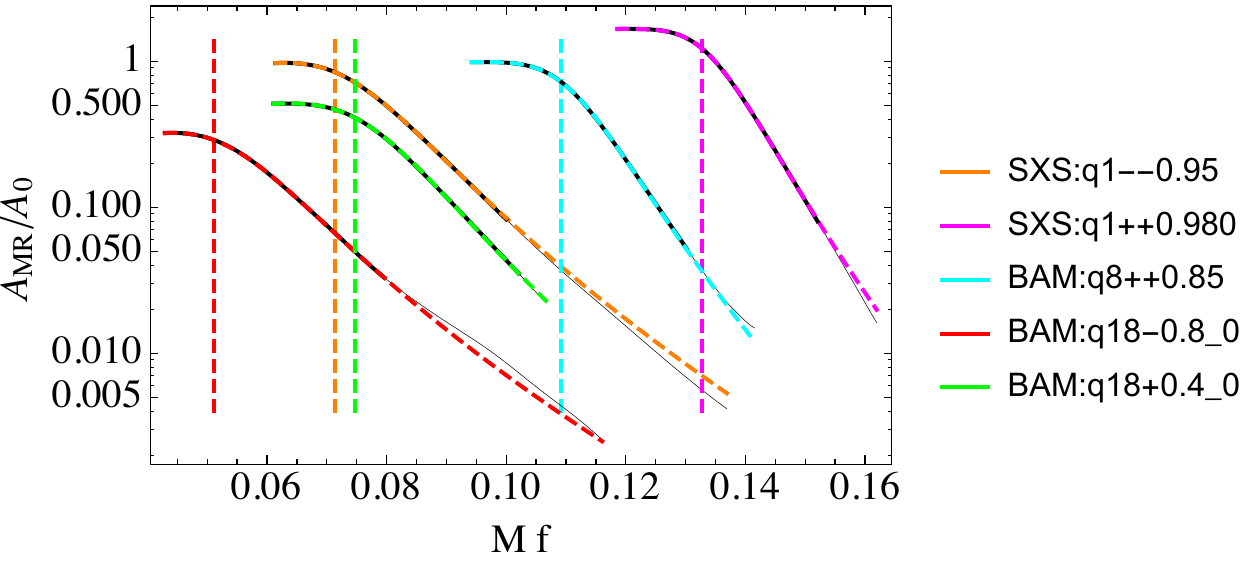}
\caption{Comparison of ringdown fit with numerical data for 5 ``corner  cases'' as indicated in the legend. Full thin lines indicate raw numerical data,  full thick lines indicate numerical data with a heuristic cutoff to remove noise at high frequencies, dashed lines indicate the best fit to ansatz Eq.~\ref{eq:AmpAMR}.
}
\label{fig:RDAmpComparison}
\end{figure}

\begin{figure}[htbp]
\includegraphics[width=\columnwidth]{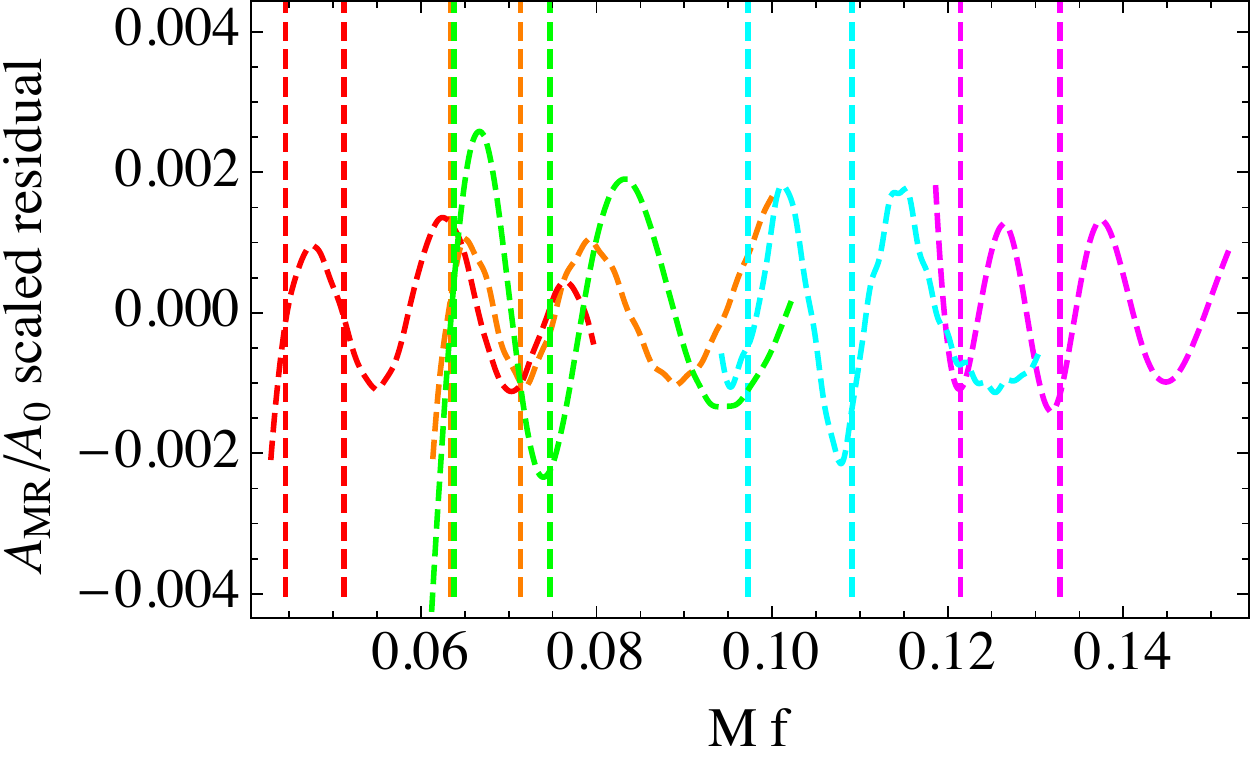}
\caption{Residuals between numerical data and best fits for the data shown in Fig.~\ref{fig:RDAmpComparison}.}
\label{fig:RDAmpResiduals}
\end{figure}

Returning now to the inspiral, it is natural to model the early frequency
behaviour as a modification of the known closed-form \PN expression
corresponding to the TaylorF2 approximant. As can be seen in Fig.\ref{fig:Amp_Corner}, modeling the pre-merger amplitude is particularly challenging for negative spins (and higher mass ratios), where the amplitude shows a very sharp drop, whereas for positive spins the frequency dependence is rather shallow. 
In the PhenomC model~\cite{Santamaria2010} the amplitude was taken directly in the form
of Eq.~(\ref{equ:xT4Amp}) as resulting from the SPA calculation.
We find however that the variation of
the phenomenological parameters is smoother if instead of using
the SPA amplitude defined in Eq.~(\ref{equ:xT4Amp}) we re-expand
the ratio of the separate \PN expansions of $A_{22}(t_f)$ and
$\ddot{\phi}(t_f)$. A comparison of different forms of the the TaylorF2 amplitude is displayed in Fig.~\ref{fig:pnAmpApproximants}
for the $q=18$ case with the larger black hole spinning at $\chi_1=-0.8$. One can see that the re-expanded forms are closer to the hybrid data. This does not hold across the parameter space, but it is for the negative spin cases like the one shown here where the difference is particularly large. 
 
\begin{figure}[htbp]
\includegraphics[width=\columnwidth]{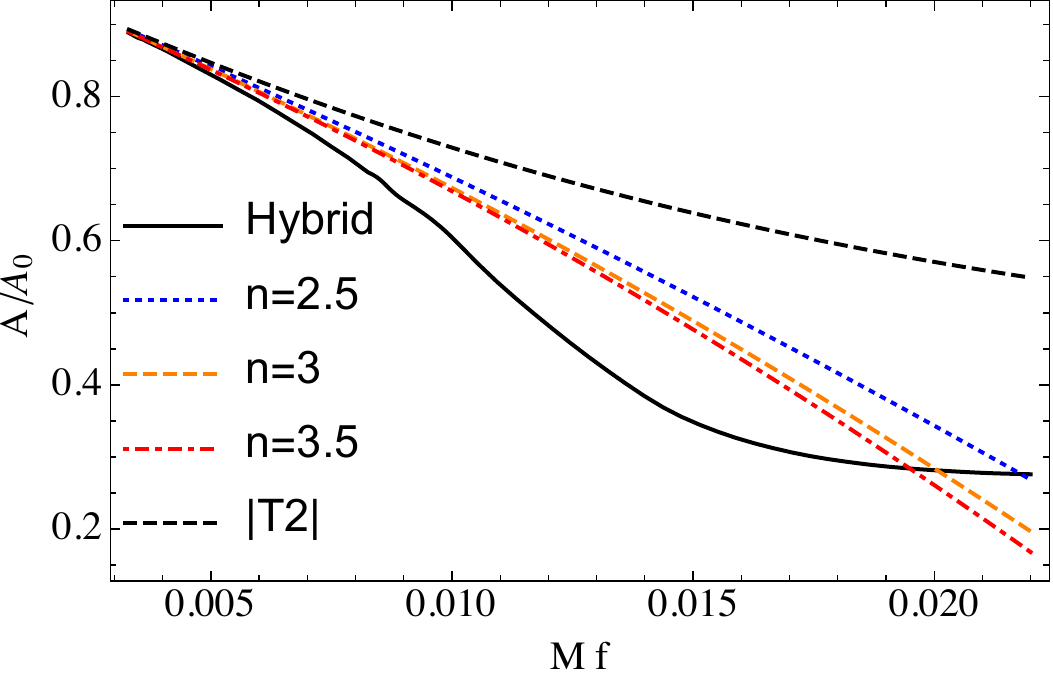}
\caption{Different forms of the TaylorF2 amplitude, rescaled as in the right panel of Fig.~\ref{fig:Amp_Corner}, are displayed for the $q=18$ case with the larger black hole spinning at $\chi_1=-0.8$. Shown are the numerical data of the hybrid waveform, the absolute value of the complex TaylorF2 amplitude from Eq.~(\ref{equ:xT4Amp}), and the real part of the TaylorF2 amplitude after re-expanding to $n=2.5, 3, 3.5$ PN order. It can be seen that the re-expanded forms are closer to the hybrid data. Taking the real part or absolute value of expressions makes no significant difference for the case and frequency range shown.}
\label{fig:pnAmpApproximants}
\end{figure} 

To perform the re-expansion we set the logarithmic terms to zero and
retain the complex terms in $A_{22}$.
Based upon comparisons to the hybrid it was only necessary to keep terms to
$\mathcal{O}(f^{2})$, which is equivalent to a consistent truncation
at 3PN order.
In the final result we use only the real part of this expression;
the contribution from the imaginary part is negligible.
The re-expanded SPA amplitude is given by,
\begin{equation}
A_{\text{PN}}(f) = A_0 \sum_{i=0}^{6} \mathcal{A}_i (\pi f)^{i/3} \, ,
\end{equation}
where $A_0$ is the leading order $f^{-7/6}$ behaviour. We then calibrate the next three natural terms in the \PN
expansion,
\begin{equation}
A_{\text{Ins}} = A_{\text{PN}}
                  + A_0  \sum_{i=1}^{3} \rho_{i} \, f^{(6+i)/3}.
\label{equ:Ains}
\end{equation}
The choice of 3 terms resulted as a compromise between accuracy and simplicity, and adding more terms may be necessary in the future to further improve the low frequency behaviour, while further progress computing post-Newtonian terms, or improvements in resumming techniques may reduce the number of calibration terms needed in the future. The frequency range for the inspiral fit was chosen by experimentation, and $fM=0.018$ was found as a good value across our parameter space of hybrids, where lowering the cutoff frequency would not significantly increase the accuracy of fits, but increasing the cutoff would increase the fit residuals, as shown in Fig.~\ref{fig:pnAmpFitResiduals} for an example case.

Since the post-Newtonian expansion shows poor convergence at higher frequencies, it is not surprising that the higher order terms we calibrate are alternating in sign, and the problem of re-constructing the signal from parameter space fits of the coefficients $\rho_i$ is poorly conditioned. We obtain significantly better results by computing the values of the fit at a set of 3 equispaced frequencies, and interpolating the values of these collocation points across the parameter space. This technique is used in the companion paper \cite{Khan2015} to construct accurate inspiral amplitude fits for the PhenomD model.

\begin{figure}[htbp]
\includegraphics[width=\columnwidth]{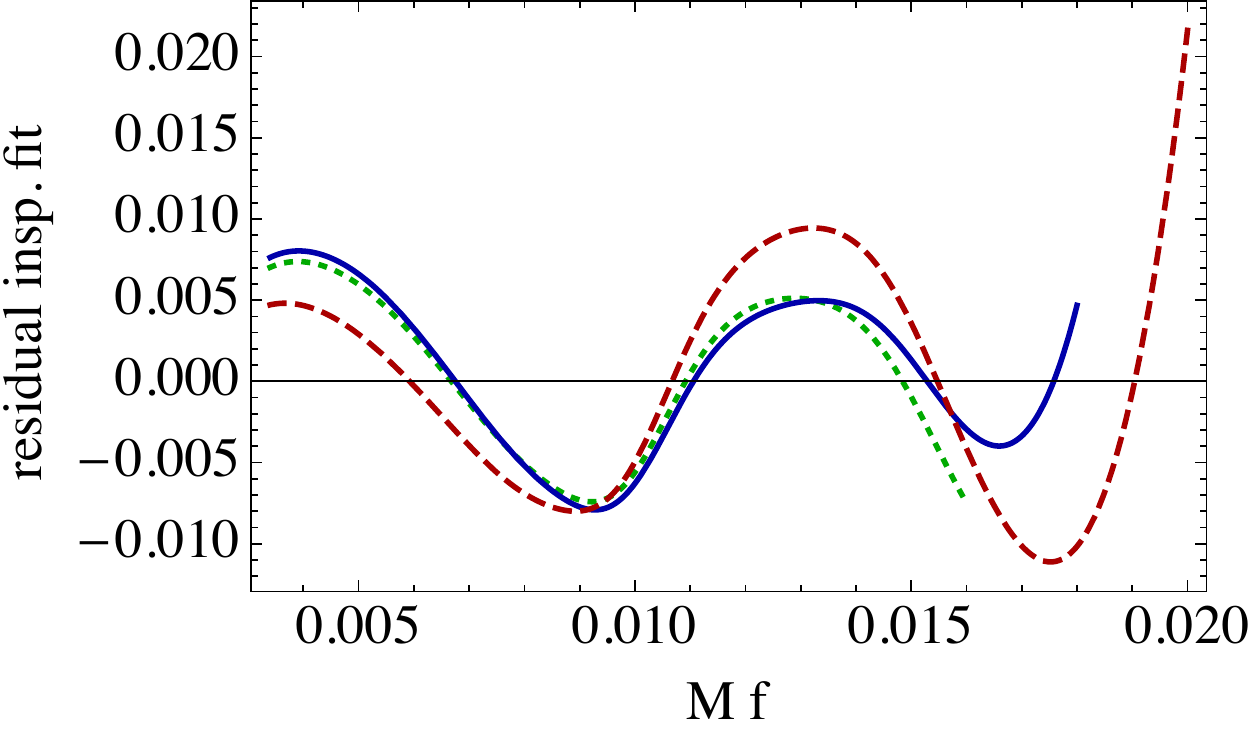}
\caption{Residuals for the inspiral fits correspding to Eq.~(\ref{equ:Ains}) with cutoff frequencies 0.016, 0.018, 0.02.}
\label{fig:pnAmpFitResiduals}
\end{figure}

\subsubsection{Phase}\label{sec:phase}

As already discussed regarding the construction of hybrid waveforms in Sec.~\ref{sec:hybrids}, waveforms from 
a system with identical intrinsic physical parameters correspond to an equivalence class of waveforms which differ by
a phase and time shift. In the time domain, both amplitude and phase are naturally affected by a time shift, in the Fourier domain the amplitude is invariant, and the phase picks up a constant and a term linear in the frequency, as can be seen by Fourier-transforming
a shifted waveform $h_{\phi_0, t_0} = h(t-t_0) e^{i  \phi_0}$:
\begin{eqnarray}\label{eq:Fourierphase_shift}
\tilde h_{\phi_0, t_0}(\omega) &=& \int_{-\infty}^{\infty} h(t-t_0) e^{i  \phi_0}  e^{-i \omega t} \, dt \nonumber\\
                                               &=& e^{i  (\phi_0+\omega t_0)} \int_{\infty}^{\infty} h(t')   e^{-i \omega t'} \, dt'.
\end{eqnarray}

Previous phenomenological waveform constructions have dealt with this ambiguity by defining a standard form of the phase function by subtracting a fit to a linear function $\phi_0 + \omega \, t_0$ over some suitably chosen frequency interval. 
In Fig.~\ref{fig:phase_alignments}, phase functions after removing different fits to  $\phi_0 + \omega \, t_0$ are shown, and the ringdown frequency $f_{\text{RD}}$ is marked. The strong dependence of the phase function on the alignment and the fact that no particular features are visible at the ringdown makes it difficult to learn about qualitative features of the phase, in particular in the merger-ringdown regime.

\begin{figure}[htbp]
\includegraphics[width=\columnwidth]{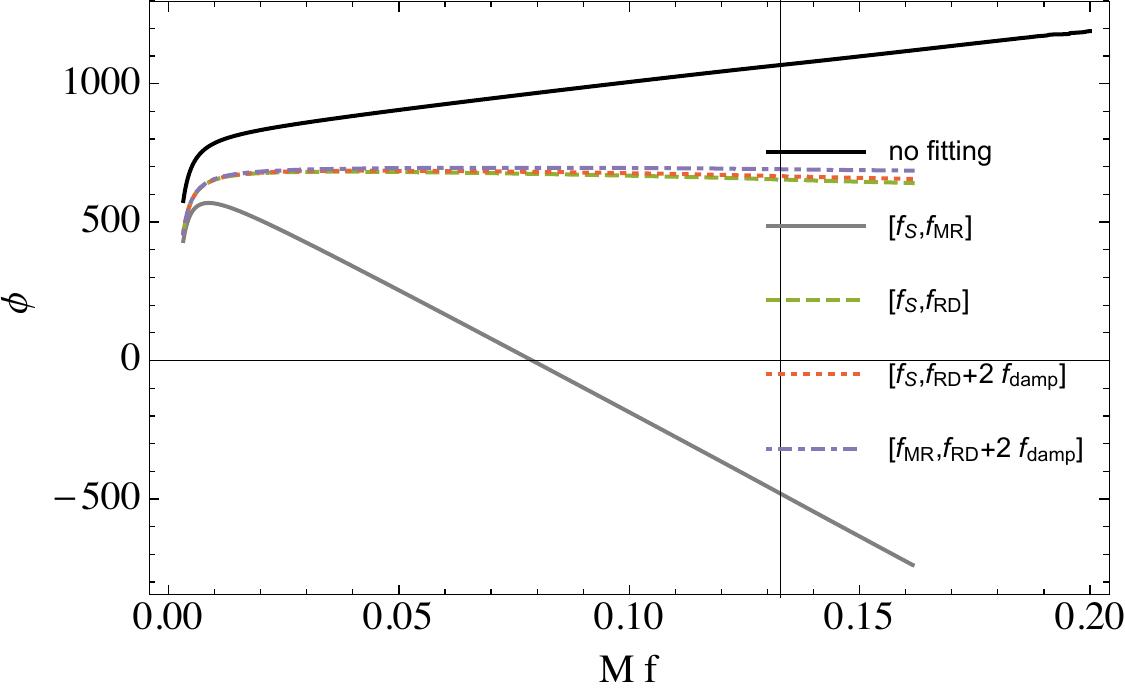}
\caption{Fourier domain phase corresponding to a hybrid constructed for SXS \NR
data corresponding to equal mass data with equal spins of $\chi_i=0.98$, with a
term $\phi_0 + \omega \, t_0$ subtracted over a frequency range. The frequency
ranges of the fits are indicated in the plot, where $M f_S = 0.0035$ is the
start frequency of clean hybrid  hybrid data, and $M f_{MR}=0.018$
the frequency where we typically separate between inspiral and merger-ringdown.}
\label{fig:phase_alignments}
\end{figure}

For this reason we take a different path to simplify the understanding of the anatomy
of the phase functions, and we consider the first and second derivatives of the
phase function with respect to the frequency. For \NR data, the
second derivative is often noisy and we thus mainly work with the first
derivative, which eliminates the phase shift ambiguity and converts the time
shift ambiguity into a simple constant offset.  In Fig.~\ref{fig:dPhase_corners}
we plot the derivative of the phase for the same 5 ``corner cases'' of maximal
mass ratio and/or spin shown in Fig.~\ref{fig:TDAmplitude_cornercases}, and mark
the ringdown frequency. We notice a characteristic structure rising above the
background and centered at the ringdown frequency, which reminds us of the
Lorentzian Eq.~(\ref{eq:defLorentzian}) we have previously employed to model the
amplitude function. Taking a derivative of the Lorentzian, we can derive that the
maxima of its derivative are located at $f_{RD}  \pm f_{damp}/\sqrt{3}$. In
Fig.~\ref{fig:ddFPhase-q2++075} we compare this prediction for the second phase
derivative with \NR data from a BAM simulation, where we can compute a
relatively clean 
second phase derivative, and we find excellent agreement. In order to capture the ``background'' in which the Lorentzian is embedded, we add terms with negative powers of the frequency $f$, to capture the steep inspiral gradient, and our basic model becomes,
\begin{equation}\label{eq:phaseMR_ansatz}
\phi'_{\text{MR}} = \alpha_{1} + \sum_{i=2}^{n} \alpha_{n} f^{-p_n}  + \frac{a}{f_{damp}^2+(f - f_{RD})^2},
\end{equation} 
where the constant $\alpha_{0}$ acts as an overall time shift of the phase function, which could be determined by matching to a description of the inspiral, and the  $\alpha_{i}$ and $a$ are free parameters which can be fit to numerical data. The number of  free parameters $\alpha_{i}$ and the choice of powers $p_n$  defines a specific merger-ringdown ansatz.
In order to find a good choice for the number of parameters and  $p_n$, we first produce fits with the free parameters $(\alpha_1, \alpha_2, p_2)$, and determine the best fit power $p_n$ across the parameter space. We find that  $p_2 \approx 2$, so we choose the integer  $p_2  = 2$ as the leading contribution. Further terms can be added to the ansatz, e.g. until the residual across the parameter space resembles pure noise, and no further information can be extracted via fits.
The specific choice made for the PhenomD model discussed in the companion paper is to add one more term with $p_3 = -1/4$.
In order to improve the convergence of Mathematica's {\tt NonlinearModelFit} function, we 
replace $f_{damp}$ by $\alpha_0 f_{damp}$ in Eq.~(\ref{eq:phaseMR_ansatz}) and also fit to $\alpha_0$.  We find that the parameter $\alpha_0$ is in the range $[0.98,1.04]$ and compensates for errors in our final mass and final spin fits near the boundary of our calibration region. For the actual PhenomD model discussed in \cite{Khan2015} this parameter will be set to unity.

For intermediate frequencies, roughly between $M f=0.018$ and half the ringdown frequency one finds that the best choice of $p_2$ for a single power law coefficient in Eq.(\ref{eq:phaseMR_ansatz}) is 
$p_2 \approx 1$, so we choose the integer  $p_2  = 1$ as the leading contribution in this regime, and we break up our model into 
 different frequency ranges in each of which we use a very simple ansatz. A more sophisticated ansatz may successfully treat the entire frequency range with a single or at least only 2 fitting regions.

For consistency with our amplitude model, we again choose $M f = 0.018$ as the transition between our inspiral and high-frequency descriptions. For the merger-ringdown part, we find that a simple choice in terms of the ringdown frequency, e.g.
$\left[ 0.45, 1.15 \right]f_{\text{RD}}$ is sufficient. The upper frequency $1.15 f_{\text{RD}}$ approximates the highest
frequency for which we have clean \NR data. For the full IMR phase we use the above fit for frequencies
larger than 0.5 $f_{\text{RD}}$.

For the intermediate phase we then choose a fitting window of $\left[ 0.018, 0.6 f_{\text{RD}}\right]$, and we use 2 fitting coefficients $p_2  = 1$, $p_3 = 4$.
                                                                                                                                                                                                                                                                                                                                                                                                                     
\begin{figure}[htbp] 
\includegraphics[width=\columnwidth]{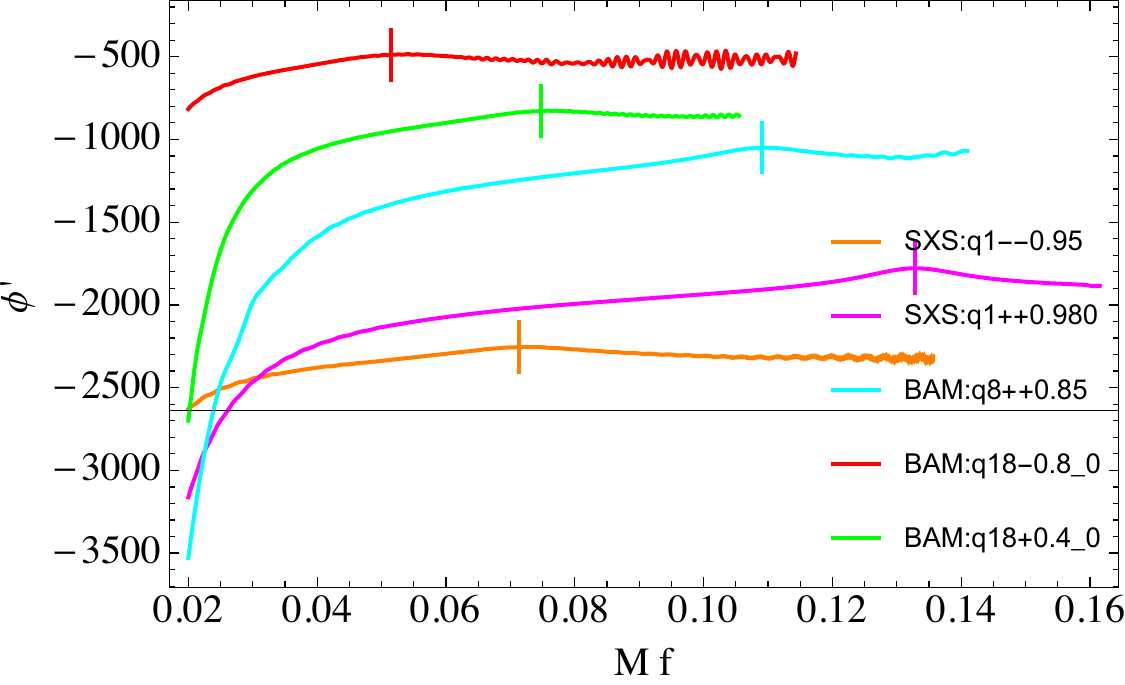} 
\caption{
Phase derivative $\phi'$ for ``corner cases'' as indicated in the legend, vertical straight lines mark the ringdown frequency $f_{RD}$.}
\label{fig:dPhase_corners}
\end{figure}

\begin{figure}[htbp]\label{fig:ddFPhase-q2++075}
\includegraphics[width=\columnwidth]{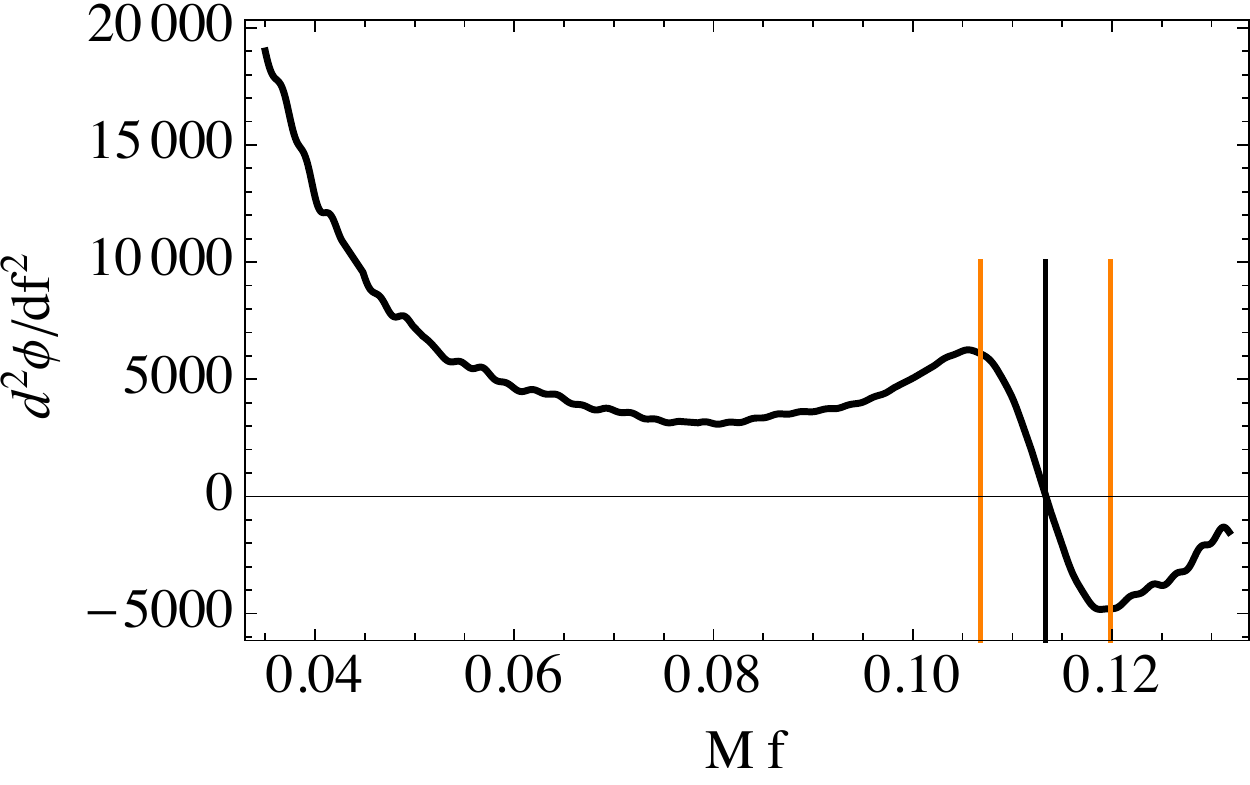} 
\caption{Second phase derivative for the BAM waveform with $q=2$, $\chi_i=0.75$, vertical straight lines mark the ringdown frequency $f_{RD}$ and $f_{RD}\pm f_{damp}/\sqrt{3}$.}
\end{figure}

We thus have succeeded in directly using information about the final spin and
mass in the ringdown description of both amplitude and phase, and turn now to
using \PN information for the inspiral. We can treat the inspiral along the
lines of the amplitude in Sec.\ref{sec:amp}. For the amplitude we have added the
next three higher order \PN terms to the 
the known TaylorF2 expression. For the phase, the 4PN terms corresponds to a
constant offset for the phase, to which we add the next three non-trivial 
higher order \PN terms, thus four \PN terms to be fitted in total.


\section{Summary} 
\label{sec:conclusions}

In this paper we have discussed preliminary work that will enable us to construct a new phenomenological waveform model for non-precessing waveforms in a companion paper \cite{Khan2015}.

We have first  presented new \NR simulations which include data
sets with a range of spin values at mass ratios 4, 8 and 18, which allow us to
 significantly extend the calibration range of current non-precessing
inspiral-merger-ringdown models. We have demonstrated sixth order convergence
for one of the $q=18$ cases and computed a phase error estimate on the order of
$0.05$ radians after aligning near merger. We find in particular small secular
drift, and errors that are dominated by noise and oscillations. We have
illustrated high accuracy also for a $\sim$10 orbit aligned-spin case, where
we have not completed the convergence series to save computational cost, and
conclude that finite difference codes based on the ``moving-puncture'' paradigm
\cite{Baker:2005vv,Campanelli:2005dd,Hannam:2006vv,Bruegmann2008} provide a very
robust tool to obtain \NR data sets in new regions of parameter space.

We have then discussed our procedure to construct hybrid \PN/numerical
relativity waveforms in the time domain, and how to convert them to the frequency
domain. Special emphasis has been put on the alignment procedure which is at the
core of the hybridization procedure, and we have presented a comparison of our
NR data and different \PN approximants, from which we conclude that
the most appropriate approximant to use in our hybrid construction
is the uncalibrated SEOBv2 model. We have also described our simple criteria to
determine an appropriate frequency range for the fitting procedure expressed in
Eq.~(\ref{eq:cleanTime}), which determines the optimal alignment of \PN and NR
waveforms. The \PN comparison has been based on a time-shift analysis directly
relevant to the hybridization procedure, which we have however also found very
useful for other applications, such as studying the time-dependent dephasing of
numerical simulations due to numerical error.

A crucial part of a waveform model is a prediction of the final mass and spin from the initial configuration, which then allows one to compute the complex quasi-normal ringdown frequencies. While there are a number of numerical fits to final mass and spin available in the literature~\cite{Barausse:2009uz,Hemberger:2013hsa,Healy:2014yta,Zlochower:2015wga}, we have presented new fits here, which extend the calibrated range, are particularly simple, and can easily be extended to new data sets or higher accuracy as further improvements of our waveform model are developed. An essential feature of the final spin fit for waveform modeling is that it respects the extreme Kerr limit. Also, the final state fits illustrate techniques for constructing parameter space fits which we will use in the companion paper \cite{Khan2015} to construct the PhenomD waveform model. 

Previous phenomenological waveform models have used a Lorentzian to model the ringdown part of the Fourier amplitude, which however leads to an incorrect high frequency falloff behaviour ($f^{-2}$ instead of $O(f^{-M}) \forall M$). Here we show that not only does a simple exponential factor lead to an excellent agreement with numerical data, but we also identify a Lorentzian of the same width, centered at the ringdown frequency in the derivative of the frequency domain phase. 

For low frequencies, we model the amplitude and phase by the standard TaylorF2
approximant, augmented by pseudo-\PN terms tuned to our hybrids
constructed from SEOBv2 and \NR data.  For the amplitude, we
use a re-expanded form with the imaginary part of the amplitude set to zero,
i.e. we neglect the very small deviations between the \GW phase and twice the
orbital phase some of us discussed in Ref.~\cite{Bustillo:2015ova}. We determine an
appropriate cutoff frequency, where our inspiral model is separated from our
merger ringdown model as $M f = 0.018$, considering the dependency of the
inspiral amplitude fit residual on this cutoff frequency.

In order to model how the phase Lorentzian is embedded in the background functional dependance of the phase, we propose to complete the model by adding inverse powers of the frequency and identify the most appropriate leading term. 

Having identified the Lorentzian feature in the phase, and a frequency up to which even a simple extension of the TaylorF2 approximant yields a good fit to the data, paves the way for complete waveform models of extremely high fidelity to the physical signal.

A first incarnation of our approach to develop phenomenological waveform models that are sufficiently accurate for the advanced detector era is the PhenomD model, which we present in a companion paper \cite{Khan2015}.


\section*{Acknowledgements} 

We thank P. Ajith and A. Taracchini for discussions.
SH, XJ, and in part AB were supported the Spanish Ministry of Economy and Competitiveness grants
FPA2010-16495, CSD2009-00064, FPA2013-41042-P, European Union FEDER funds, Conselleria d'Economia i Competitivitat del Govern de les Illes Balears and Fons Social Europeu.
MH was supported by the Science and Technology Facilities Council grants ST/H008438/1
and ST/I001085/1, and FO and MP by ST/I001085/1.
SK was supported by Science and Technology Facilities Council.

{\tt BAM} simulations were carried out at Advanced Research Computing 
(ARCCA) at Cardiff, as part of the European PRACE petascale computing
initiative on the clusters Hermit, Curie and SuperMUC, and on the UK DiRAC Datacentric cluster.


\bibliography{PhenomPaper1}

\end{document}